\documentclass[twocolumn,appendixfloats,trackchanges]{aastex6}
\turnoffedit

\begin{document}


\title{A Suzaku View of Accretion Powered X-ray Pulsar GX~1+4}


\author{Yuki Yoshida\altaffilmark{1,2}, Shunji Kitamoto\altaffilmark{1,2}, Hiroo Suzuki\altaffilmark{1}, Akio Hoshino\altaffilmark{1,2}, Sachindra Naik\altaffilmark{3}}
\and
\author{Gaurava K. Jaisawal\altaffilmark{3}}

\altaffiltext{1}{Department of Physics, College of Science, Rikkyo University, 3-34-1 Nishi-Ikebukuro, Toshima, Tokyo 171-8501, Japan}
\altaffiltext{2}{Research Center for Measurement in Advanced Science, Rikkyo University, 3-34-1 Nishi-Ikebukuro, Toshima, Tokyo 171-8501, Japan}
\altaffiltext{3}{Astronomy and Astrophysics Division, Physical Research Laboratory, Navrangapura, Ahmedabad - 380009, Gujarat, India}
\email{yy@rikkyo.ac.jp}


\begin{abstract}
We present results obtained from a {\it Suzaku} observation of the accretion powered X-ray pulsar GX~1+4. 
\edit1{Broad-band continuum spectrum of the pulsar was found to be better described by a simple model consisting of a blackbody component and an exponential cutoff power-law than the previously used \texttt{compTT} continuum model.}
Though the pulse profile had a sharp dip in soft X-rays ($<$10 keV), phase-resolved spectroscopy confirmed that the dimming was not due to increase in photoelectric absorption. 
Phase-sliced spectral analysis showed the presence of a significant spectral modulation beyond 10 keV except for the dip phase. 
A search for the presence of cyclotron resonance scattering feature in the {\it Suzaku} spectra yielded a negative result. 
Iron K-shell (K$_\alpha$ and K$_\beta$) emission lines from nearly neutral iron ions ($<$Fe~I\hspace{-.1em}I\hspace{-.1em}I) were clearly detected in the source spectrum. 
A significant K$_\alpha$ emission line from almost neutral Ni atoms was detected for the first time in this source. We estimated the iron abundance of $\sim$80\% of the solar value and Ni/Fe abundance ratio of about two times of the solar value. 
We searched for a iron Ly$_\alpha$ emission line and found a significant improvement in the spectral fitting by inclusion of this line. 
We found a clear intensity modulation of the iron K$_\alpha$ line with the pulse phase with an amplitude of 7\%. 
This finding favored an inhomogeneous fluorescent region with a radius of much smaller than the size ($\sim$3$\times$10$^{12}$ cm) estimated by an assumption of homogeneous matter.
\end{abstract}


\section{INTRODUCTION} \label{sec:introduction}
GX~1+4 is a peculiar accretion powered X-ray pulsar with a long pulse period of about 150~s \citep{Doty1976}. 
Early observations in 1970s showed the pulsar to be very bright in X-rays and exhibiting regular spin-up ($\nu/\dot{\nu}=0.02\,\rm{yr}^{-1}$) \citep{Nagase1989} which is in good agreement with the standard accretion torque model (e.g. \citealt{Ghosh1979a,Ghosh1979b}). 
However, in 1980s, the luminosity of the pulsar decreased by at least two orders of magnitude during which it remained undetectable \citep{HallDavelaar1983}. 
When the pulsar reappeared, it showed a spin-down activity \citep{Makishima1988}, suggesting the occurrence of torque reversal event. 
According to the standard accretion torque model \citep{Ghosh1979b}, the observed torque reversal in GX~1+4 indicates the surface magnetic field of the neutron star to be $\sim$10$^{14}$~G \citep{Makishima1988, Dotani1989, Mony1991}. 
As this value is extremely high, an alternative model e.g. a retrograde disk model was also discussed by \citet{Makishima1988} and \citet{Dotani1989}. \citet{Chakrabarty1997} suggested that the formation of a retrograde disk denied the need for an unusually strong magnetic field and naturally explained the anti-correlation between torque and luminosity in 20--60~keV band as observed with {\it BATSE}. 
A marginal detection of cyclotron resonance scattering features at $\sim$34~keV \citep{Rea2005,Naik2005,Ferrigno2007} indicated the magnetic field of the neutron star to be of the order of 10$^{12}$~G. 
Therefore, the question of the magnetic field strength of GX~1+4 is still open. 
The companion star of GX~1+4 is a late type giant star \citep{GlassFeast1973,Davidsen1976}. 
Optical counterpart of the neutron star was classified as an M5 I\hspace{-.1em}I\hspace{-.1em}I spectral type giant star in a rare type of symbiotic system \citep{Shahbaz1996,ChakrabartyRoche1997}. 
\edit1{Based on variations of the pulse period of the neutron star during spin-up phase measured with the high-energy X-ray spectrometer onboard OSO-8, \citet{Cutler1986} proposed an orbital period of approximately 304 d. 
\citet{Pereira1999} claimed confirmation of this period using {\it BATSE} observation in 1990s when the source was in spin-down phase. 
On the other hand, using infrared measurement of radial velocity of the M giant, the orbital period was derived to be 1161~d by \citet{Hinkle2006}. 
X-ray light curves of GX~1+4 did not show any modulation either at the 1161~d orbital period derived from infrared observations or at previously reported 304~d period from OSO-8 observations \citep{Corbet2008}. }
Although the distance to GX~1+4 has a large uncertainty \citep{ChakrabartyRoche1997}, we assume the distance to source to be 4.3~kpc \citep{Hinkle2006} in this paper. The pulse profile of GX~1+4 has been reported to have a characteristic shape and a prominent dip in medium- and low- intensity state \citep{Dotani1989,Galloway2001,Kotani1999,Naik2005}.
Such a sharp dip in pulse profile is also seen in several other sources; GX~304-1 (e.g. \citealt{McClintock1977}), 4U~1626-67 (e.g. \citealt{Kii1986}), A0535+262 (e.g. \citealt{Mihara1995}), RX~J0812.4-3114 (e.g. \citealt{Reig1999}) and so on. 
This sharp dip in GX~1+4 is interpreted as due to the eclipse of the X-ray emitting region by the accretion column of the pulsar \citep{Dotani1989,Galloway2000}. This argument was based on the evidence of increase in the scattering optical depth accompanied with the sharp dip \citep{Galloway2000}. 
A detailed analysis of the individual dips in GX~1+4 suggested that the width of the dips is proportional to the source flux \citep{Galloway2001}. 
This indicates that the features of this dip change with the source luminosity and this in fact will give us a clue to understand the accretion flow geometry at different mass accretion rates.

The energy spectrum of accretion powered X-ray pulsar is often represented by a model consisting of a 
power-law with high energy cutoff, known to be the signature of unsaturated Comptonization, and Gaussian 
functions for the iron emission lines. In intermediate and high luminosity states of GX~1+4 
($F_{2\mbox{\scriptsize --}10{\rm keV}}=10^{-10}$--$10^{-9}\,\rm{erg\,s^{-1}\,cm^{-2}}$), 
the source spectrum has been described by either cutoff power-law model or an analytical 
model based on thermal Comptonization of hot plasma close to the source (\texttt{compTT} 
in XSPEC) \citep{Galloway2000,Galloway2001,Naik2005, Ferrigno2007}. \citet{Galloway2000} 
proposed that the \texttt{compTT} model reproduced the observed spectrum of GX~1+4 with 
scattering taking place in the accretion column. Phase-sliced spectroscopy of GX~1+4 by 
using \texttt{compTT} model provided us insight of the information on the accretion flow 
and the accretion column geometry \citep{Galloway2000,Galloway2001,Naik2005,Ferrigno2007}.


It is well known that GX~1+4 exhibits bright iron K-shell emission lines 
\citep{Kotani1999,Dotani1989, Naik2005,Paul2005}. The long pulsation period 
of the pulsar makes it appropriate to investigate the phase dependence of emission 
line properties. \citet{Kotani1999} analyzed K-shell emission lines from lowly ionized 
iron ions by using {\it Ginga} and {\it ASCA} observations of the pulsar and found a 
positive correlation between the equivalent width and the absorption column density of 
the circumstellar matter. Using parameters such as ionization state of the iron ions 
(Fe\,I--I\hspace{-.1em}V), absorption column density and estimated X-ray luminosity, 
\citet{Kotani1999} suggested that the line emitting region in GX~1+4 consisted of lowly 
ionized plasma and extended up to 10$^{12}$ cm from the neutron star.

Along with iron K$_{\alpha}$ line, a strong K$_{\beta}$ emission line with equivalent 
width of $\sim$550~eV was also detected during an extended low state of the pulsar in 
2000 \citep{Naik2005}. On the other hand, \citet{Paul2005} reported a discovery of a 
Ly$_{\alpha}$ emission line in the absence of K$_{\beta}$ line in the pulsar spectrum. 
They suggested that the diffuse gas in the Alfv\'en sphere and/or accretion curtains 
to magnetic poles are the possible iron Ly$_{\alpha}$ line emitting regions in GX~1+4.
The distance and size of the line emitting region can be determined by examining the 
intensity modulation of emission lines with respect to the neutron star rotation.
Considering the detection of emission lines corresponding to different ionization 
states at different luminosity levels, it is important to carry out detailed spectral 
investigation of the pulsar by using data from detectors with good energy resolution 
to have a clear understanding of the matter distribution in the binary system. 

{\it Suzaku} observation of GX~1+4 provided us an opportunity to analyze the 
phase-sliced broad-band spectra and emission line diagnostics because of its 
high sensitive detectors with very good energy resolution as well as the long 
spin period of the pulsar. In this paper, we report mainly the results obtained 
from the detailed spectral analysis of {\it Suzaku} observation of the pulsar. 
These results will help in understanding the emission mechanism around the 
neutron star and the origin of the emission lines.

\section{OBSERVATION AND DATA REDUCTION} \label{sec:obs}

\subsection{Suzaku Observation} \label{subsec:suzakuobs}
{\it Suzaku} observation of the pulsar GX~1+4 (OBSID $=$ 405077010) was carried out 
from 2010 October 02 06:43 (UT) to October 04 12:20 (UT). During the observation, the 
X-ray Imaging Spectrometers (XISs; \citealt{Koyama2007}) were operated in normal mode 
incorporating a 1/4 window option which ensures a time resolution of 2~s. Spaced-row 
Charge Injection (SCI) was also performed with 2~keV equivalent-electrons for XIS~0 
and XIS~3 (front-illuminated or FI CCDs) and 6~keV equivalent-electrons for XIS~1 
(back-illuminated or BI CCD). The Hard X-ray Detector (HXD; \citealt{Takahashi2007}) 
was operated in the standard mode wherein individual events were recorded with a time 
resolution of 61~$\mu$s. The target was placed at the HXD nominal position. Though the 
total on-source duration during the observation was $\sim$192 ks, the effective exposures 
with XIS-0, XIS-1, XIS-3, HXD/PIN and HXD/GSO were 97.3 ks, 99.7 ks, 99.7 ks, 88.3 ks and 
82.2 ks, respectively.

\subsection{Data Reduction} \label{subsec:suzakuspec}
{\it Suzaku} archival data of GX~1+4 were analyzed by using HEASARC software version 6.16--6.18. 
On-source events of the XISs were extracted from a circular region of 3' radius with center at the source position. Background events were extracted from the XIS data by selecting an annulus region (from 4' to 5') away from the source with the pulsar co-ordinates as the center. 
\edit1{A pile-up estimation following \citet{Yamada2012} showed a maximum pile-up of 1\% around the source position. 
Therefore we neglected the effect of the pile-up. }
A quick look analysis of the XIS~1 data showed an apparently inconsistent energy-scale compared to those of XIS~0 and XIS~3. 
This inconsistency has been interpreted as the Self Charge Filling effect (SCF effect; \citealt{Todoroki2012}). 
Events for XIS~0 and XIS~3 were well recovered from the degradation of the charge transfer efficiency with the SCI though it was insufficient for XIS~1. 
We corrected XIS~1 data from the SCF effect by applying the method proposed by \citet{Todoroki2012} (see Appendix). 
The redistribution matrix files and ancillary response files for XISs were generated by using \texttt{xisrmfgen} and \texttt{xissimarfgen} routines \citep{Ishisaki2007}, respectively. 
In our analysis, response file released in 2010 July was used for HXD/PIN data whereas response and effective area files released in 2010 May were used for HXD/GSO data.

\begin{figure}
\begin{center}
  \includegraphics[width=75mm]{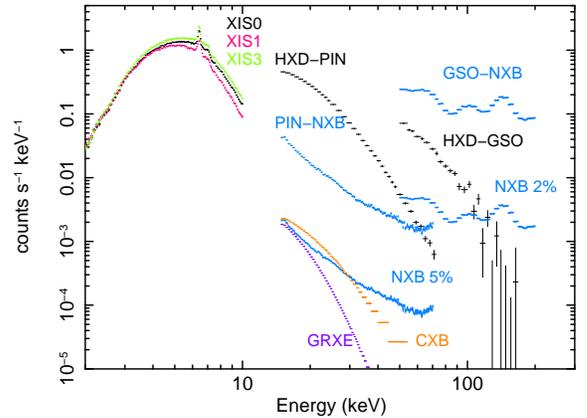}
\end{center}
\caption{Background-subtracted phase-averaged spectra of GX~1+4 obtained 
from XISs (below 10 keV) are shown in black, red and green colors, whereas 
NXB and CXB subtracted HXD spectra (above 15~keV) are shown in black color. 
The expected NXB spectra and their repeatability are plotted with blue 
colors and the CXB and GRXE are shown with orange and purple colors.}
\label{fig:gxPhAveSpec}
\end{figure}

Background subtracted spectra obtained from the {\it Suzaku} observation of GX~1+4 are shown in 
Figure~\ref{fig:gxPhAveSpec}. In the same figure, expected background spectra for the HXD/PIN and 
HXD/GSO are also plotted. We used tuned (LCFITDT) Non-X-ray Background (NXB) models \citep{Fukazawa2009} 
to get the NXB events for HXD/PIN and the HXD/GSO. Repeatability of the NXB for HXD/PIN is about 5\%. 
Therefore, it is at most 5\% of the signal from GX~1+4 even in the energy range around 50 keV. After 
NXB subtraction from the HXD/PIN data, we also subtracted an expected contribution of Cosmic X-ray 
Background (CXB) by  using a typical model by \citet{Boldt1987}. Since CXB amounts to only $\sim\,5\%$ 
of NXB, its ambiguity due to the sky-to-sky fluctuation ($\sim11\,\% \,\rm{at}\,1\sigma$) corresponds to 
at most $\sim\,0.6\%$ of NXB and is negligibly small.

As GX~1+4 is located at Galactic coordinates of ($\mathit{l}$,$\mathit{b}$)=(1\fdg9\,,\,4\fdg8), Galactic Ridge X-ray Emission (GRXE) may contaminate the data as well. {\it INTEGRAL}/IBIS mapping observations showed that the typical 17--60~keV GRXE flux is less than \edit1{$2\times10^{-9}\,\rm{erg\,s^{-1}\,cm^{-2}\,FOV^{-1}}$ }in the Galactic bulge region $|\mathit{l}|<10^{\circ}$(Figure~13 of \citealt{Krivonos2007}). 
However, within the HXD/PIN field of view (FOV), we expected the contribution of GRXE flux of $\lesssim\,2 \times10^{-11}\,\rm{erg\,s^{-1}\,cm^{-2}}$, and it  amounts to only $\sim\,3\%$ of NXB.
Therefore, we eventually subtracted only the NXB and the CXB from the HXD/PIN data. From HXD/GSO data, 
we subtracted only the NXB as the contributions from CXB and GRXE are negligible. 
From a careful investigation of {\it INTEGRAL} observation, we did not find any contaminating X-ray source in the FOV of HXD/PIN during {\it Suzaku} observation of GX~1+4 whereas many X-ray sources were present in the FOV of HXD/GSO. 
Using data from {\it INTEGRAL}/IGRIS observation, we examined the contribution of contamination for the source. Since all the contaminating sources were near the edge of the FOV of HXD/GSO, with a small effective area of HXD/GSO \citep{Matsumoto1999}, the expected count rate from those sources was only $\lesssim$\,2\% of the source in the energy range of the HXD/GSO and hence we neglected them.

\section{ANALYSIS AND RESULTS}\label{sec:result}

\subsection{Timing Analysis}\label{subsec:timing}
\begin{figure}
\begin{center}
  \includegraphics[width=75mm]{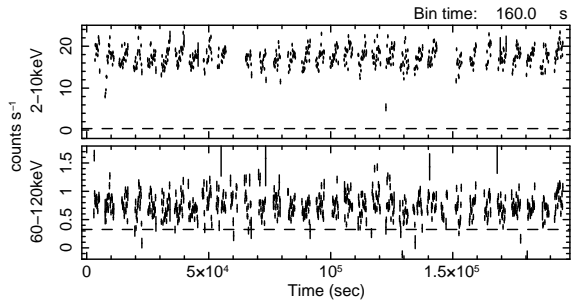}
\end{center}  
  \caption{
  Background-subtracted, 2--10 keV (XIS) and 15--60 keV (HXD/PIN) light curves of 
  GX~1+4 with 160~s time bins are shown in top and bottom panels, respectively. 
  Data from all three XISs are added together. Dashed-lines in both panels represent 
  the background levels in each energy band.}
  \label{fig:gxlc}
\end{figure}

For timing analysis, we applied barycentric correction to the arrival times of individual photons using the \texttt{aebarycen} task of FTOOLS. 
Light curves with time resolutions of 2~s and 1~s were extracted from XIS (2--10~keV) and HXD/PIN (15--60~keV) event data, respectively. 
\edit1{Figure~\ref{fig:gxlc} shows the background-subtracted light curves with a bin-time of 160~s i.e. at the pulsar spin period, in soft (top panel) and hard X-rays (bottom panel). }
Although there are several data points with low count rate, there is no trend of any gradual intensity change during the {\it Suzaku} observation of the pulsar.

\subsubsection{Pulse Profile}\label{subsubsec:PulseProfile}

\begin{figure}
\begin{center}
  \includegraphics[height=75mm]{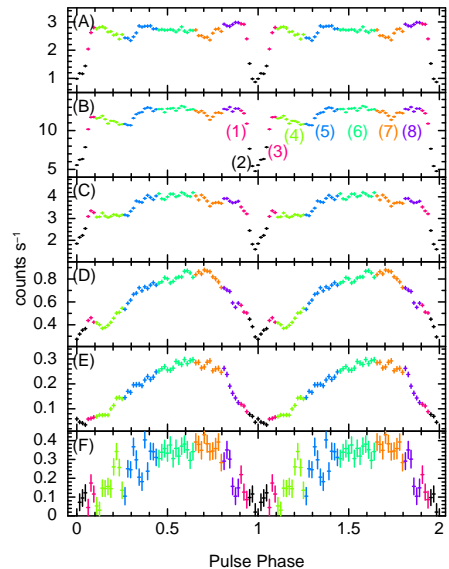}
\end{center}
\caption{Energy resolved pulse profiles of GX~1+4 obtained from background-subtracted XISs 
(data added from three sensors), HXD/PIN and HXD/GSO light curves in 2.0--4.0~keV (A), 
4.0--7.0~keV (B), 7.0--10.0~keV (C), 15.0--25.0~keV (D), 25.0--60.0~keV (E) and 60--114~keV (F) 
energy ranges. For phase-resolved spectroscopy, the data in each panel are divided into eight 
pulse phase intervals as indicated by quoted numbers and different colors.}
\label{fig:gxEnedivPulseProf}
\end{figure}

We searched for pulsations in the light curves obtained from XISs (2--10~keV range), HXD/PIN 
(15--60~keV range) and HXD/GSO (60--114~keV range) by using the standard epoch folding technique 
\citep{Leahy1983}. Our analysis revealed a consistent barycentric pulsation period of 
$P=159.9445\pm0.0002$~s at an epoch of 55471.3 MJD, in all the light curves. Our estimated 
pulse period agrees well with the expected value derived by considering the earlier measurements 
of period and period derivative of the pulsar \citep{Gonzalez2012}. Using the estimated pulse 
period and time of intensity minimum (55471.2796 MJD) as epoch, we generated pulse profiles of 
the pulsar by applying \texttt{efold} task of FTOOLS. Pulse profiles obtained from background 
subtracted light curves in 2.0--4.0 keV, 4.0--7.0 keV, 7.0--10.0 keV (data from all three XISs 
added together), 15.0--25.0 keV, 25.0--60.0 keV (HXD/PIN) and 60.0-114 keV (HXD/GSO) ranges are 
shown in top to bottom panels of Figure~\ref{fig:gxEnedivPulseProf}, respectively. The HXD/PIN 
and HXD/GSO data have been corrected for dead time. Gradual change in the shape of the energy 
resolved pulse profiles can be clearly seen in the figure. The sharp dip in the soft X-ray pulse 
profiles was prominent though there was no spike-like feature in the dip as reported by 
\citet{Dotani1989}. The width of the dip was found to be increasing with energy as pointed 
out by \citet{Naik2005}. Apart from the dip, a small hump was also seen at phase $\sim$0.1 in 
15.0--25.0 keV range pulse profile.

\subsection{Phase-Averaged Spectroscopy}\label{subsec:PhAveSpec}

\subsubsection{Continuum of Broad-band Spectrum}\label{subsubsec:PhAveCont}
In this section, we describe phase-averaged spectroscopy of GX~1+4 by using XISs, HXD/PIN 
and HXD/GSO data obtained from the {\it Suzaku} observation of the pulsar. As shown in 
Figure~\ref{fig:gxPhAveSpec}, prominent iron K-shell emission lines and significant absorption 
in the low energy band are evident even without any spectral model fitting. Since the XIS 
spectra show strong absorption, the detected events below 2~keV are dominated by the ``low 
energy tail'' component \citep{Matsumoto2006} which is characteristic of the instruments.
\citet{Suchy2012} reported that this ``low energy tail'' component had not been well calibrated 
due to which there is a small miss-match between the FI and BI CCDs. Therefore, we used data 
from FI CCDs in 2--10 keV range in our broad-band spectral fitting.

\begin{figure}
\begin{center}
  \includegraphics[width=75mm]{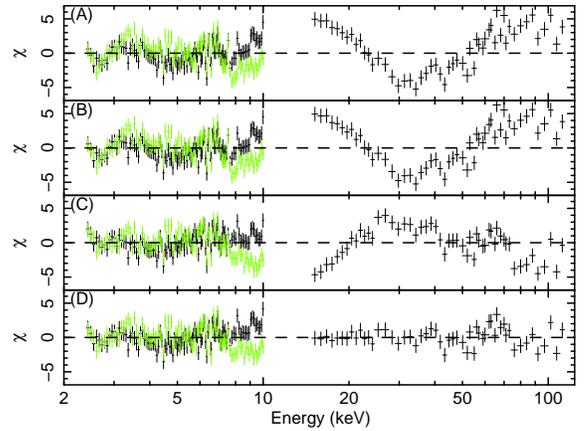}
\end{center}
\caption{
\edit1{Residuals divided by its error}, which were obtained from the spectral fitting of XIS, HXD/PIN and HXD/GSO data of the {\it Suzaku} observation of GX~1+4. 
\edit1{The residuals divided by its error}, obtained from fitting the 2-120 keV range spectra using (A) \texttt{compTT} model for disk geometry, (B) \texttt{compTT} model for spherical geometry, (C) a cutoff power-law model and (D) a blackbody $+$ cutoff power-law (BB+CPL) model are shown in panels, respectively.
}
\label{fig:gxFitResultPhaseAve}
\end{figure}

Broad-band (2--120 keV range) spectra of GX~1+4 were simultaneously fitted with the earlier reported \texttt{compTT} continuum model along with the photoelectric absorption (\texttt{TBabs} in XSPEC) by matter along the line of sight \citep{Galloway2000}. 
We applied both the geometries e.g. a disk and a sphere by using the analytical approximation at a fixed red-shift of 0. 
The residuals obtained from the spectral fitting by assuming disk and spherical geometries are shown 
in panel (A) and (B) of Figure~\ref{fig:gxFitResultPhaseAve}, respectively. 
The best-fit parameters obtained from these fittings are column density of photoelectric absorption $N_{\rm H}=1.1\times10^{23}$atoms cm$^{-2}$, photon source temperature $T_{\rm o}=1.4~{\rm keV}$, hot electron temperature $T_{\rm e}=11~{\rm keV}$, normalization of the \texttt{compTT} model component $A_{\rm{c}}=1.8\times 10^{-2}$ and optical depths for disk geometry $\tau_{\rm disk}=4.4$ and spherical geometry $\tau_{\rm sphere}=9.6$. 
These values are nearly consistent with those of earlier reported values obtained from the {\it RXTE} \citep{Galloway2000}, {\it Beppo}SAX \citep{Naik2005} and {\it INTEGRAL} \citep{Ferrigno2007} observations. 
However, presence of global wavy structures in the residuals can be clearly seen in the panels (A) \& (B) of Figure~\ref{fig:gxFitResultPhaseAve}, yielding poor values of reduced $\chi^2$s ($\chi^{2}_{\nu}=4.4$ for 261 $d.o.f.$) for both the geometries.

As an alternative representation of the continuum, we attempted our spectral fitting by using an empirical model consisting of an exponential cutoff power-law continuum (CPL) along with photoelectric absorption (\texttt{TBabs}). 
This simple model fitted the 2--120 keV range spectrum better than the \texttt{compTT} model yielding reduced $\chi^{2}$ of 2.9 ($d.o.f.=262$). 
However, the residuals obtained from this fitting still showed wavy structures (panel (C) of Figure~\ref{fig:gxFitResultPhaseAve}). 
\edit1{Addition of a blackbody (BB) component to the exponential cutoff power-law continuum model, improved the spectral fitting further. 
The reason for better spectral fitting with the addition of blackbody component is as follows. 
The BB+CPL model has an additional free parameter compared to the \texttt{compTT} model and can determine the slope of the high-energy part (index of the CPL) and the cutoff energy independently from the low energy range of the spectrum. 
On the other hand, the slope and the cutoff energy of the \texttt{compTT} model are not independent at low-energy range.}
Though the value of the reduced $\chi^{2}$ obtained from fitting the data with BB+CPL model is still poor ($\chi^{2}_{\nu}/d.o.f.=2.1/260$), 
the wavy structures are now absent in the residuals (panel (D) of Figure~\ref{fig:gxFitResultPhaseAve}). 
Therefore, we used BB+CPL model as the most suitable model to describe the broad-band (2--120 keV range) spectrum of GX~1+4. 
A possible reason for the poor value of reduced $\chi^2$ can be due to the spectral variations at different pulse phases as shown in Figure~\ref{fig:gxEnedivPulseProf}. 
\edit1{The values of parameters obtained from spectral fitting with BB+CPL model are column density of photoelectric absorption $N_{\rm H}=(1.30\pm0.02)
\times10^{23}$~H\,atoms\,cm$^{-2}$, blackbody temperature $kT_{\rm BB}=1.66\pm0.03$~keV, radius of blackbody emitting region $R=0.63^{+0.16}_{-0.20}$~km (assuming the distance of 4.3~kpc), photon index $\Gamma=0.46^{+0.03}_{-0.04}$ and cutoff energy $E_{\rm cutoff}=22.1\pm0.6$~keV. }
Using this model, the unabsorbed source flux in 1--120 keV range was estimated to be $F_{1\mbox{\scriptsize --}120{\rm keV}}=4.1\times10^{-9}\,{\rm erg\,s^{-1}\,cm^{-2}}$.

\subsubsection{Emission lines and related features}\label{subsubsec:PhAveLine}
As discussed in previous section, the broad-band continuum spectrum of GX~1+4 can be better described by a model consisting of a blackbody component and a cutoff power-law component (BB+CPL) along with interstellar absorption. 
As seen in Figure~\ref{fig:gxPhAveSpec}, presence of several emission lines and other features are distinctly visible even without spectral fitting. 
For a detailed analysis of emission lines and related features in GX~1+4, we restricted our spectral fitting to 5.8--7.8 keV energy range. In this restricted region, the continuum can be fitted by a simple power-law (PL) model with an absorption edge so that we can deduce the edge parameter independently from the low energy absorption. 
However, the large value of photoelectric absorption with $N_{\rm{H}}$=1.30$\times10^{23}$~H\,atoms\,cm$^{-2}$ affects the intensities of the emission 
lines. 
Thus, we applied the photoelectric absorption (\texttt{TBabs}) to only the Gaussian functions used for emission lines by fixing it at above value. Note that, for our broad-band spectroscopy, we used only the FI CCDs because of a discrepancy between the FI and BI CCDs in low energy region (below 3~keV). 
However, in the above restricted narrow energy range, the discrepancy between the FI and BI CCDs is negligibly small. Therefore, we used data from all three CCDs for the analysis of emission lines and related features in GX~1+4.
 
\begin{figure*}
\gridline{\fig{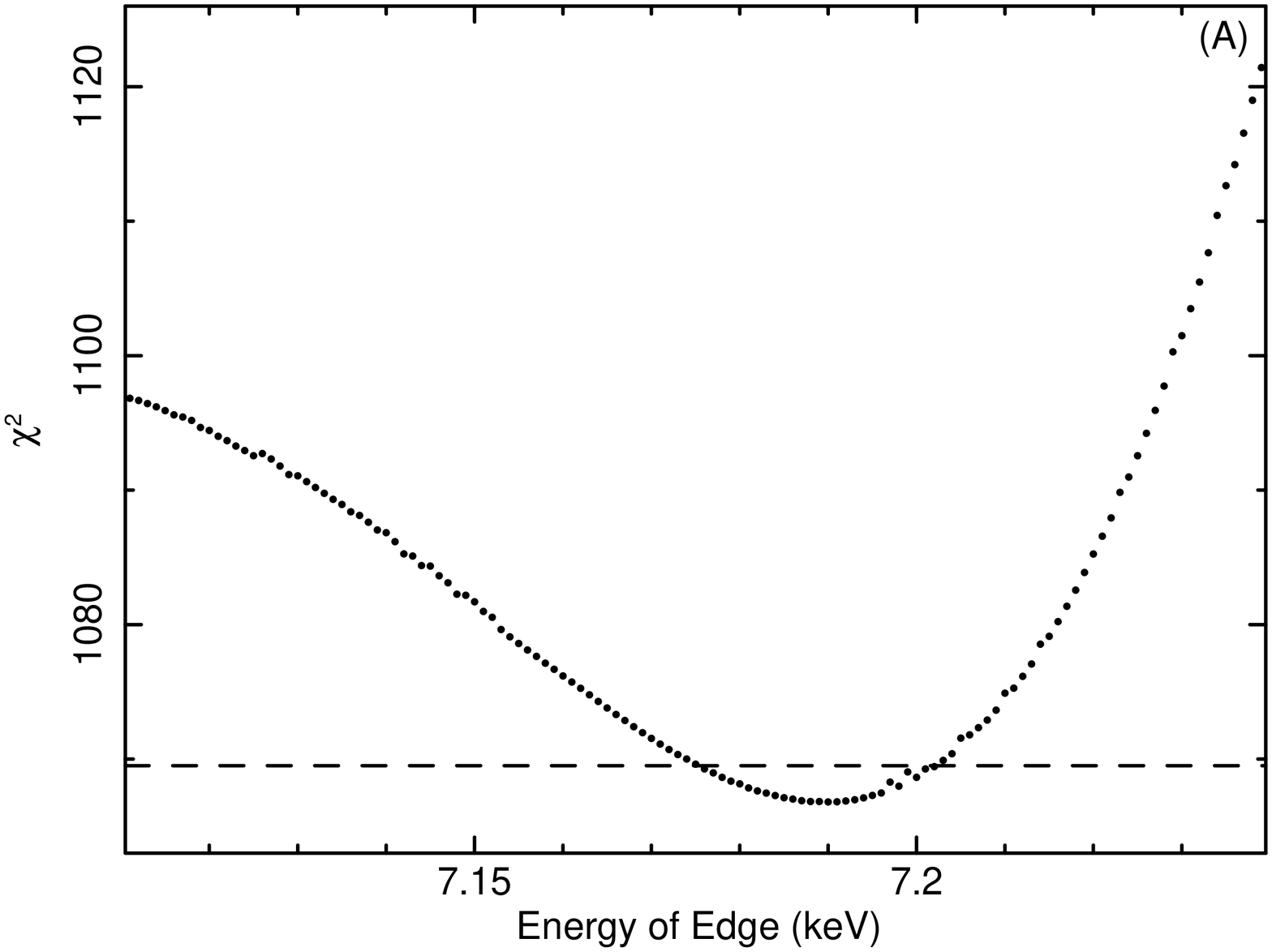}{0.4\textwidth}{}
	     \fig{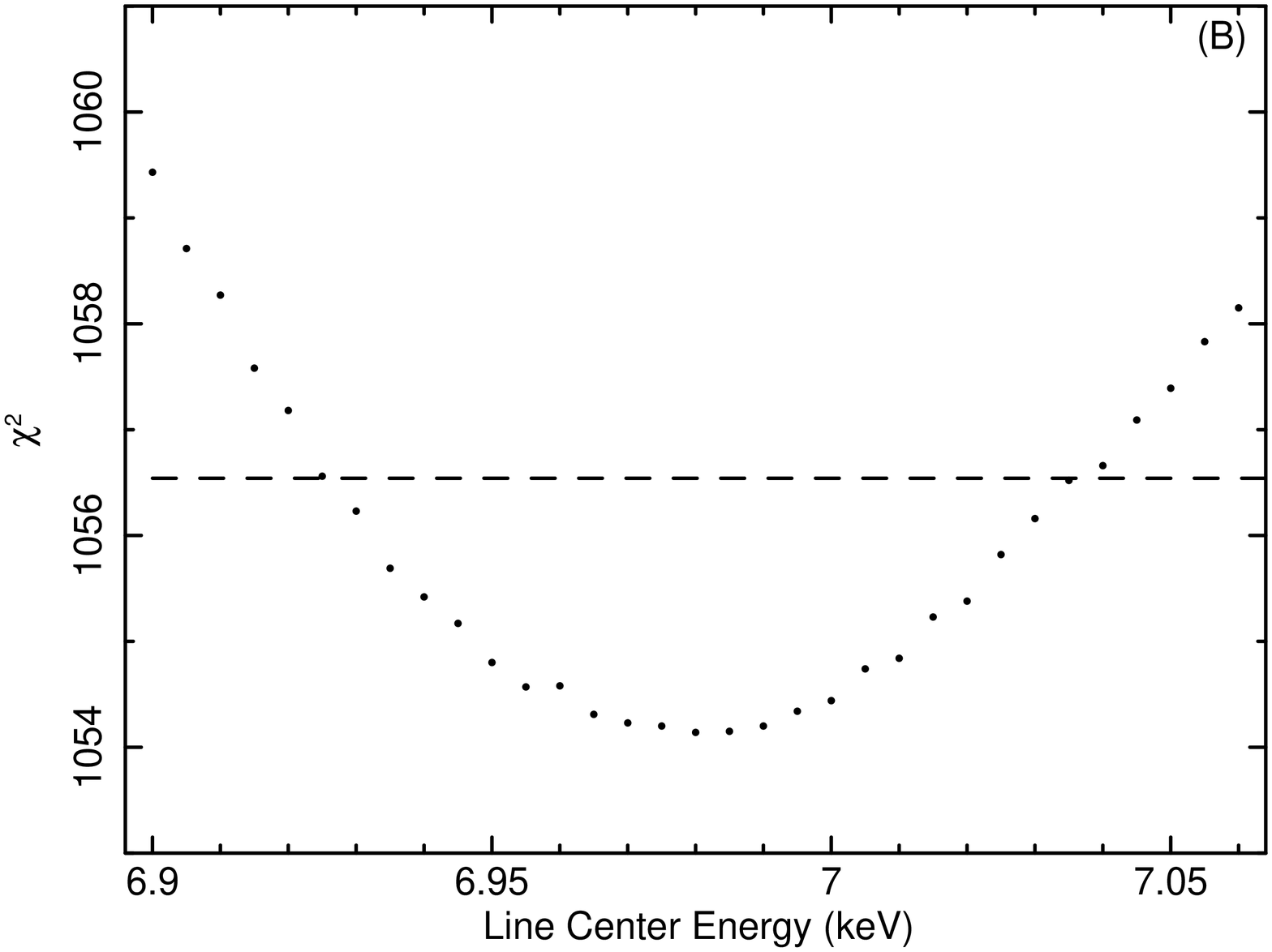}{0.4\textwidth}{}}
  \caption{
  (A) The resultant $\chi^2$ values as a function of energy of iron absorption edge, obtained from the fitting with Model\,A in Table\,\ref{tab:PhaseAveLineParam} with 
  the number of degrees of freedom being 966. 
  (B) The resultant $\chi^2$ values as a function of energy of iron Ly$_{\alpha}$ line, obtained from the fitting with Model\,B in Table\,\ref{tab:PhaseAveLineParam} with the number of degree of freedom being 965. For each panel, dashed lines represent 90~\% confidence level.}
\label{fig:gxEdgeChi}
\end{figure*}

Since the intense emission line feature at around 6.4~keV is known to be the iron K$_{\alpha}$ line, 
the iron K$_{\beta}$ line should appear at around 7.1 keV. 
As the iron K$_{\beta}$ line contaminates the absorption edge feature, it is difficult to determine the parameters corresponding to these structures. 
\edit1{As the ionization state of the iron is low for 6.4~keV K$_{\alpha}$ emission line, we fixed the ratio of the energy of the K$_{\beta}$ to that of the K$_{\alpha}$ lines to 1.103, which is for the neutral case \citep{Yamaguchi2014}, in the fitting. }
We also assumed the line to be sufficiently narrow. 
Considering these, we fitted the 5.8--7.8 keV data obtained from three XIS sensors by a model consisting of a PL continuum with two Gaussian functions representing iron K$_{\alpha}$ and K$_{\beta}$ emission lines, taking into account of the photoelectric absorption of 1.30$\times$10$^{23}$~H\,atoms\,cm$^{-2}$. 
Apart from these, an absorption edge component was added to the model. 
The edge energy was scanned from 7.10~keV to 7.25~keV with a step of 1~eV. 
The resultant $\chi^2$ values as a function of the edge energy is plotted in the left panel of Figure~\ref{fig:gxEdgeChi}. 
The best values of energy of absorption edge and K$_{\alpha}$ emission line are $7.19\pm0.01$~keV and $6.425\pm0.001$~keV, respectively. 
The resultant energy ratio of the absorption edge and the K$_{\alpha}$ line is $1.119\pm0.002$. 
This ratio corresponds to that of Fe\,I\hspace{-.1em}I \citep{Lotz1968} which is consistent to the energy ratio of K$_{\beta}$ and K$_{\alpha}$ emission lines. 
The intensity ratio of iron K$_{\alpha}$ and K$_{\beta}$ emission lines is estimated to be 0.11$\pm$0.02 which is also consistent to that of Fe\,I--I\hspace{-.1em}I\hspace{-.1em}I\citep{Yamaguchi2014}.

\begin{figure}
\begin{center}
   \includegraphics[width=75mm]{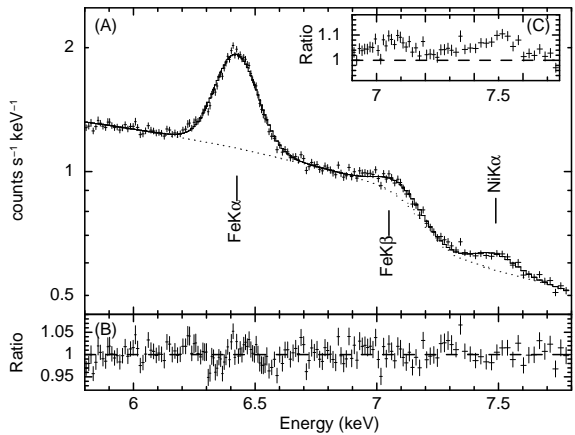}
\end{center}
   \caption{
   (A) Phase-averaged spectrum in 5.8--7.8~keV energy range fitted with a model consisting of a power-law continuum, an absorption edge and three Gaussian components for emission lines. 
    The photoelectric absorption (\texttt{TBabs}) is multiplied only to three Gaussian components. 
   (B) Ratio of the data to the best fit model in panel (A). 
   (C) Ratio of data to a model which is produced by removing the components of the iron K$_{\beta}$ and Ni K$_{\alpha}$ lines from the best fit model.}
   \label{fig:gxLineFitResultPhaseAve}
\end{figure}

In the residuals obtained from our spectral fitting, we found excess at around 7.5~keV. 
Addition of another narrow emission line component in the model reduced the $\chi^{2}$ from 1166.51 $(d.o.f.=968)$ to 1066.89 $(d.o.f.=966)$. This improvement corresponds to a very high statistical significance (the null hypothesis probability is much less than 10$^{-10}$) through $F$-test. 
The energy of the third emission line was determined to be $7.49\pm0.02$~keV which is identified as K$_{\alpha}$ line from neutral or lowly ionized Ni. 
The best-fit parameters obtained from our analysis are summarized in Table~\ref{tab:PhaseAveLineParam} (Model~A). 
The data and the best-fit model are shown in Figure~\ref{fig:gxLineFitResultPhaseAve}(A) whereas Figure~\ref{fig:gxLineFitResultPhaseAve}(B) shows the ratio of the data to the best fit model. 
For a demonstration, the ratio to the best fit model without iron K$_{\beta}$ and Ni K$_{\alpha}$ emission lines are shown in Figure~\ref{fig:gxLineFitResultPhaseAve}(C). 
\edit1{This discovery of Ni K$_{\alpha}$ line in GX~1+4 is mainly due to the better energy resolution and the good statistics of {\it Suzaku}/XIS compared to the earlier detectors.}

\citet{Paul2005} reported the detection of iron K$_\alpha$ and Ly$_\alpha$ emission lines in the absence of K$_\beta$ emission line in GX~1+4 with the High Energy Transmission Grating Spectrometer (HETGS) onboard {\it Chandra}. 
We also searched for an iron Ly$_\alpha$ line in the XIS spectra of the pulsar. 
Before adding the iron Ly$_\alpha$ line component to our model, we fitted the phase averaged spectra by keeping the width of K$_\beta$ line as a free parameter. 
The values obtained are summarized in Table~\ref{tab:PhaseAveLineParam} (Model~A'). 
We found a significant broadening of the K$_\beta$ line with $100\pm30$~eV, in the standard deviation of the Gaussian function with 90\% confidence level. 
The null hypothesis probability of increasing freedom as being free of the width was $1.2\times10^{-3}$ using $F$-test. 
\edit1{This broadening of the K$_\beta$ line suggests that emission line feature is contaminated with other features, such as a Ly$_\alpha$ line.}
We added another line component at around 6.9 keV as iron Ly$_\alpha$ line to the model consisting of a PL continuum, an absorption edge and three Gaussian functions for emission lines and taking into account of a photoelectric absorption (\texttt{TBabs}). 
\edit1{The fitting was conducted with scanning the center energy of iron Ly$_\alpha$ line from 6.9~keV to 7.06~keV with a step of 5~eV. 
Then the energy of edge was fixed at 7.190~keV, which was obtained in the analysis of emission lines and related features without iron Ly$_\alpha$ component.
In the fitting, the lines were assumed to be sufficiently narrow.}

Figure~\ref{fig:gxEdgeChi}(B) shows the resultant $\chi^2$ values as a function of \edit1{the center energy of iron Ly$_{\alpha}$ emission line.}
Addition of iron Ly$_\alpha$ line component to the model significantly improved the fitting with value of $\chi^{2}$ decreased from 1066.89 ($d.o.f.=966$) to 1054.14 ($d.o.f.=965$). 
The null hypothesis probability of the inclusion of this line component was estimated to be $6.9\times10^{-4}$. 
The results obtained from our analysis are listed in Table~\ref{tab:PhaseAveLineParam} (Model~B). 
\edit1{The line center energy and intensity, estimated from {\it Suzaku} observation of the pulsar were found to be $6.98\pm0.06$~keV and $(0.8^{+0.2}_{-0.3})\times10^{-4}$ photons\,s$^{-1}$\,cm$^{-2}$, respectively, which are consistent with that reported by \citet{Paul2005}. 
If the iron Ly$_\alpha$ line component was included in the model, the resultant intensity ratio of iron K$_\alpha$ and K$_\beta$ emission lines was $0.10\pm0.02$, which is still consistent to that of Fe\,I-I\hspace{-.1em}I\hspace{-.1em}I within their errors.}

\edit1{We also searched for a He-like iron K$_{\alpha}$ emission by fixing its center energy to be 6.70~keV, which is expected from the best-fit value of iron Ly$_{\alpha}$ emission line. }
We did not find any improvement in the value of $\chi^{2}$ with the addition of the Gaussian function. 
The 90~\% upper limit of the line intensity was estimated to be $N_{\rm Fe\,He_{\alpha}}=1\times10^{-5}$ photons\,s$^{-1}$\,cm$^{-2}$. 
The best-fit parameters are listed in Table~\ref{tab:PhaseAveLineParam} (Model~C).

\subsubsection{Search for an absorption feature at around 34~keV}\label{subsubsec:AbsorptionFeature}
We searched for a cyclotron absorption line at around 34~keV as reported earlier from {\it Beppo}SAX \citep{Rea2005, Naik2005} and {\it INTEGRAL} \citep{Ferrigno2007} observations of the pulsar although there was no visual indication of any such feature in the fit residuals of our phase-averaged spectra. 
In the beginning, the phase-averaged spectra obtained from HXD/PIN and HXD/GSO were fitted with the addition of an absorption component -- a multiplicative absorption edge or Gaussian absorption line (\texttt{gabs} in XSPEC; $\exp[(-\tau_{\ast}/\sqrt{2\pi}\sigma_{\rm{w}}) \exp(-(E-E_{\rm abs})^{2}/2\sigma_{\rm{w}}^{2} )]$), where the width of the Gaussian absorption line, $\sigma_{\rm{w}}$, was fixed at 4~keV as in \citet{Ferrigno2007}. 
The absorption feature at $\sim$34~keV could not be detected as its significance was less than 3$\sigma$. The 90\% confidence upper limit for maximum absorption depth $\tau_{\ast}$ of the line was $<0.7$. 
Further, we searched for an absorption feature between 30~keV and 110~keV by using a Gaussian absorption line model, the width of which was fixed at 5~keV, 
10~keV and 15~keV. Consequently, we conclude the absence of any detectable absorption feature in above energy range.

\subsection{Phase-Sliced Spectroscopy}\label{subsec:PhaseSlicedSpec}
Pulse profile of the pulsar, as shown in Figure~\ref{fig:gxEnedivPulseProf}, was found 
to be characterized by several energy dependent features like the sharp and prominent 
dip with variable width, hump-like feature etc. This prompted us to carry out phase-sliced 
spectral analysis of the {\it Suzaku} observation of GX~1+4. For this, we divided data into 
eight pulse phase bins as shown in Figure~\ref{fig:gxEnedivPulseProf} with different colors. 
Phase intervals 1 and 3 (marked with red color in the figure) were combined into one. The 
phase interval 2 (marked with black color in the figure) is around the intensity minimum 
phase and we call this phase ``dip interval''.

\begin{figure}
\begin{center}
\includegraphics[width=75mm]{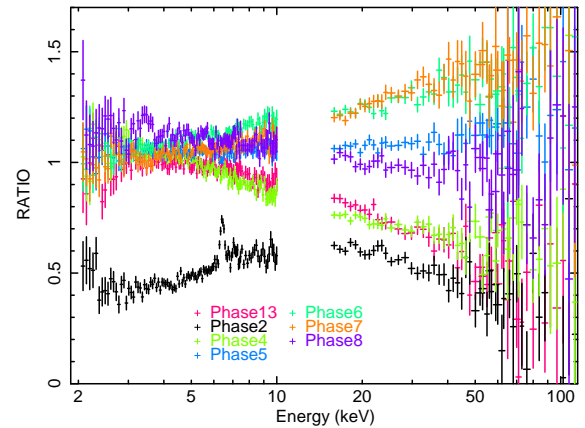}
\end{center}
\caption{Spectral ratio between the phase-sliced and phase-averaged spectra of GX~1+4. 
Different colors correspond to the colors indicated in Figure~\ref{fig:gxEnedivPulseProf}.  }
\label{fig:gxPhDivSpecHR}
\end{figure}

In order to grasp the basic properties of the pulsar among all the phase-sliced spectra in a 
model-independent way, we first calculated spectral ratios e.g. dividing the phase-sliced 
spectra by the phase-averaged spectrum. The resulted spectral ratios are shown in 
Figure~\ref{fig:gxPhDivSpecHR}. The modulation below $\sim$7~keV was found to be small for 
all the phase-sliced spectra except the dip interval. The spectral ratio for the dip interval 
did not only show a simple absorption feature but implied the absence of a spectral component 
that mainly contributes in the energy range below 10~keV. In hard X-ray ranges, increase in the 
amplitude of spectral modulation with energy suggested the change in power-law photon index for 
different phase bins. Apart from the spectral changes with pulse phase of the pulsar, large 
equivalent width of the iron K$_{\alpha}$ emission line can be clearly seen in the dip interval. 

\begin{figure*}[h]
\gridline{\fig{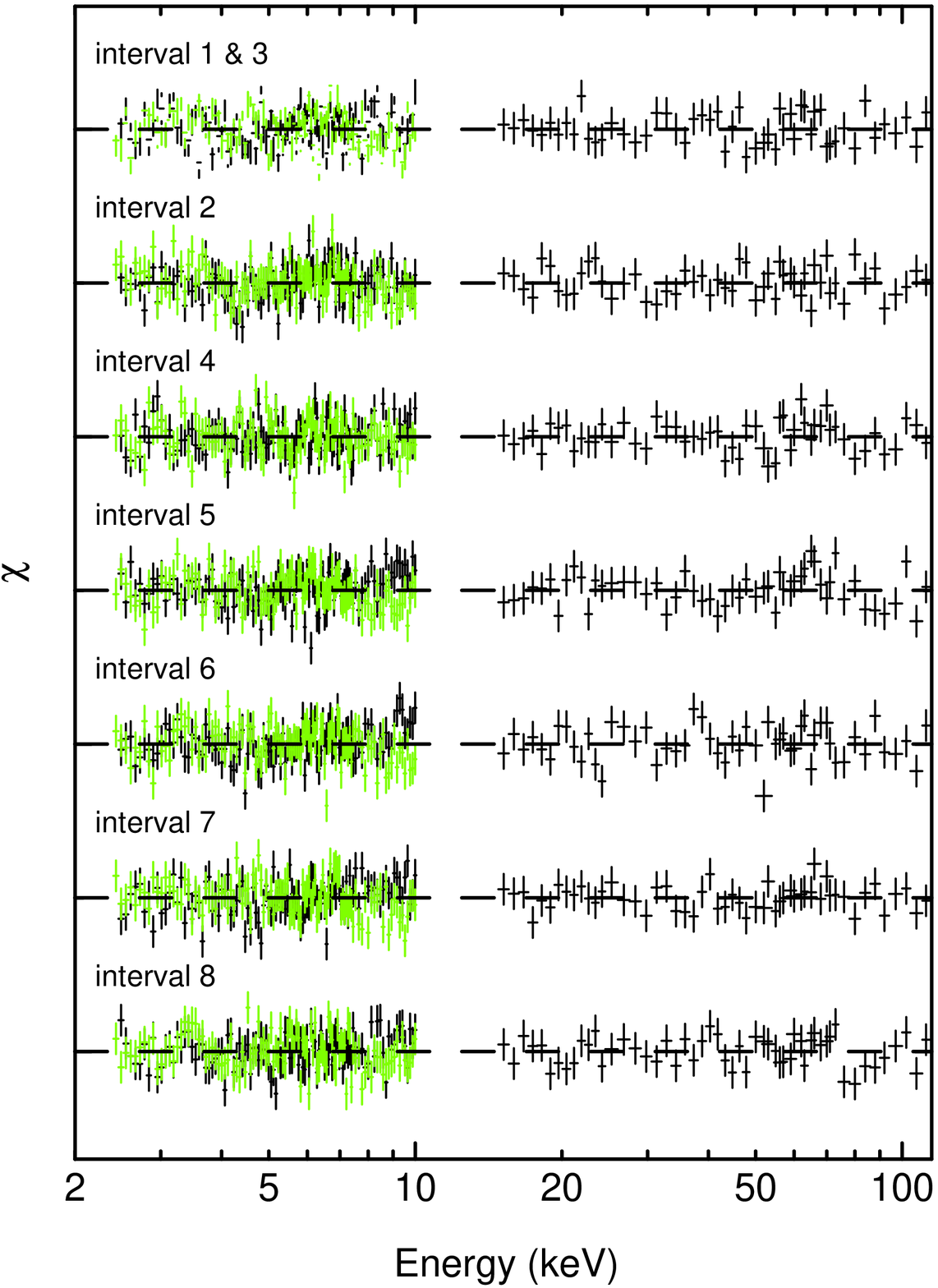}{0.3\textwidth}{}
  \fig{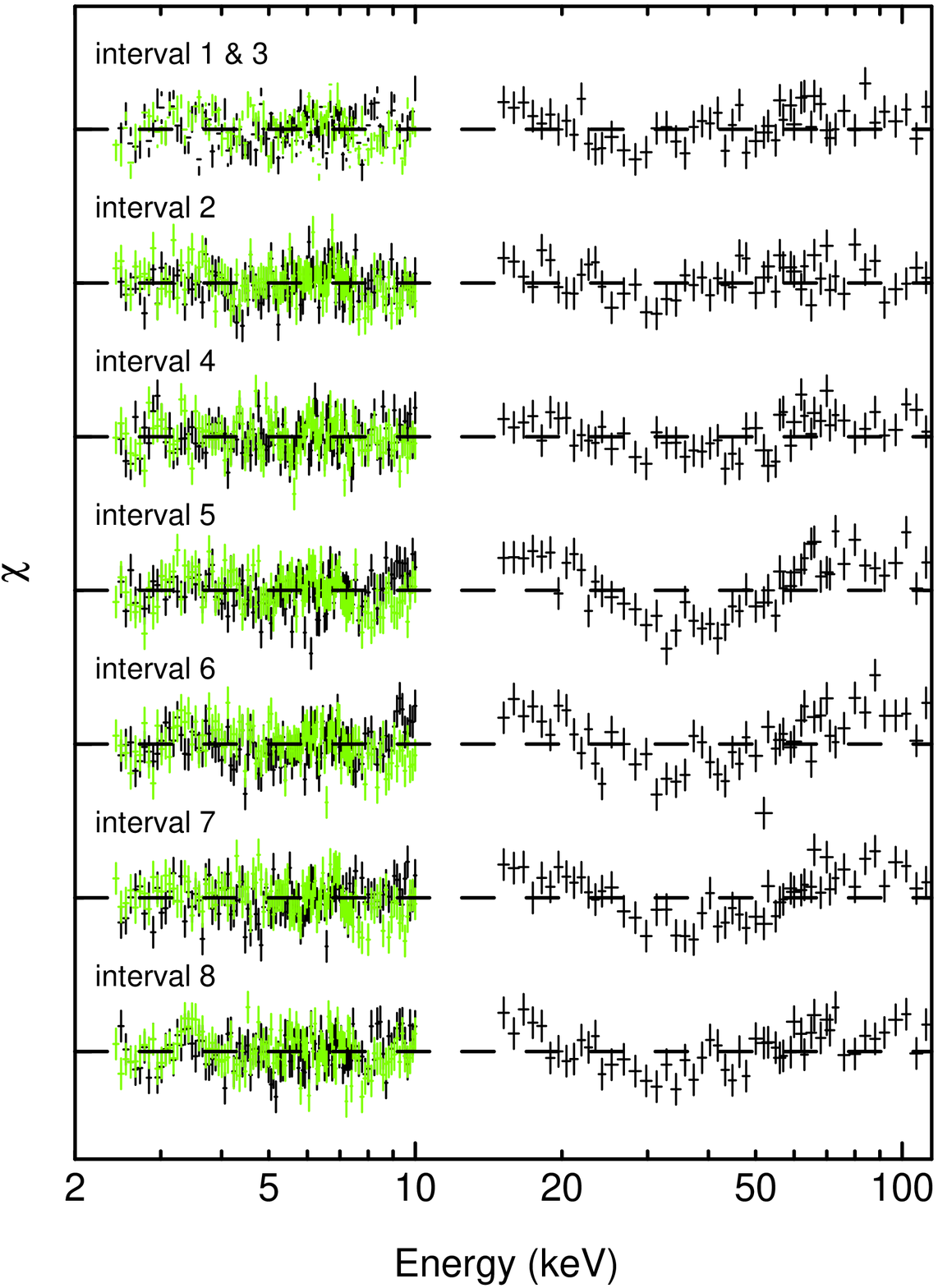}{0.3\textwidth}{}
  \fig{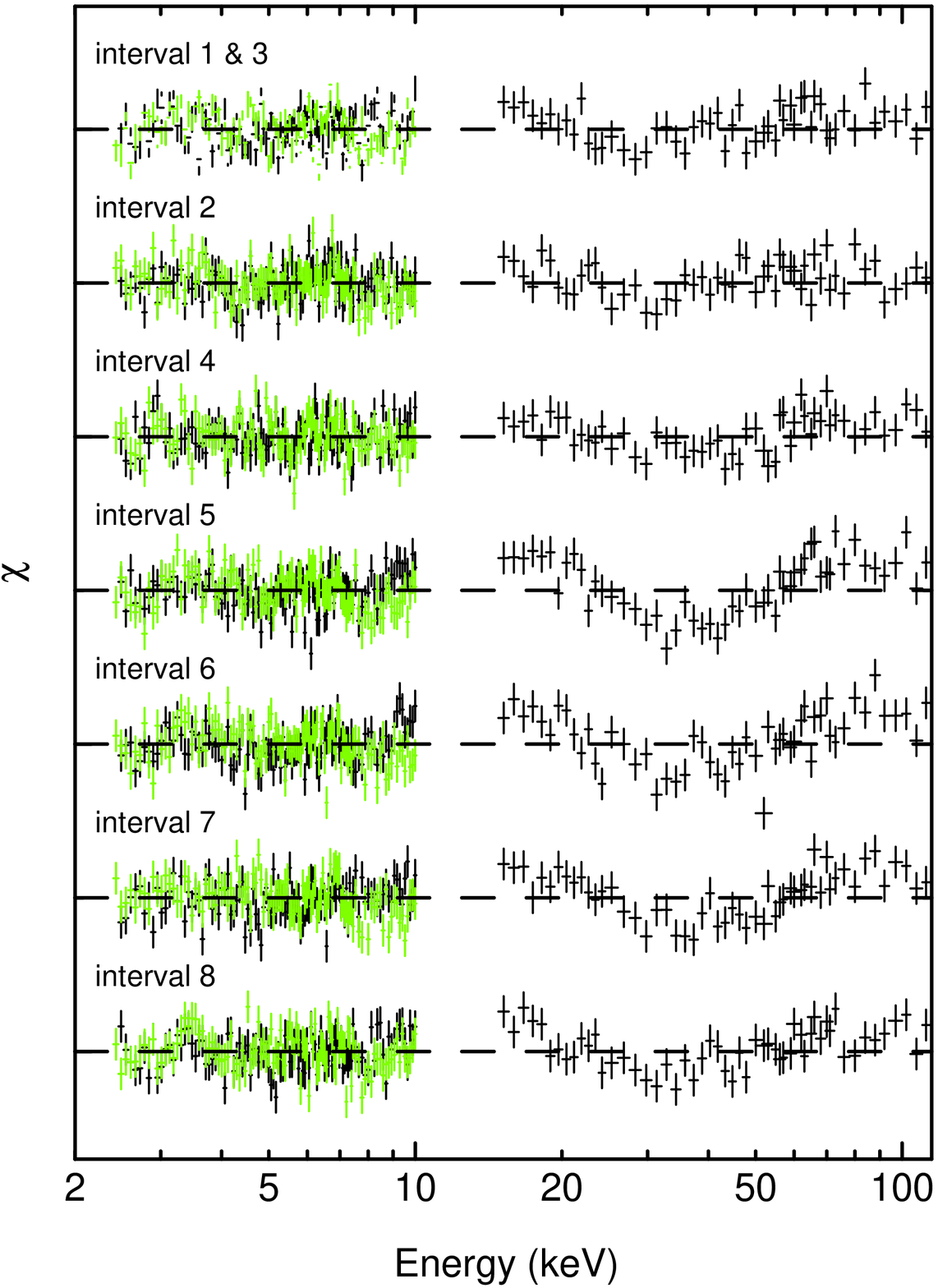}{0.3\textwidth}{}}
\caption{\edit1{Residuals divided by its error} of the spectral fitting of phase-sliced spectra of GX~1+4 with BB+CPL model 
(left panels), \texttt{compTT} model with a disk geometry (middle panels) and \texttt{compTT} 
with a spherical geometry (right panels). Dashed horizontal lines indicate the zero axis.}
\label{fig:gxFitResultPhaseDiv}
\end{figure*}

\begin{figure*}
\gridline{	\fig{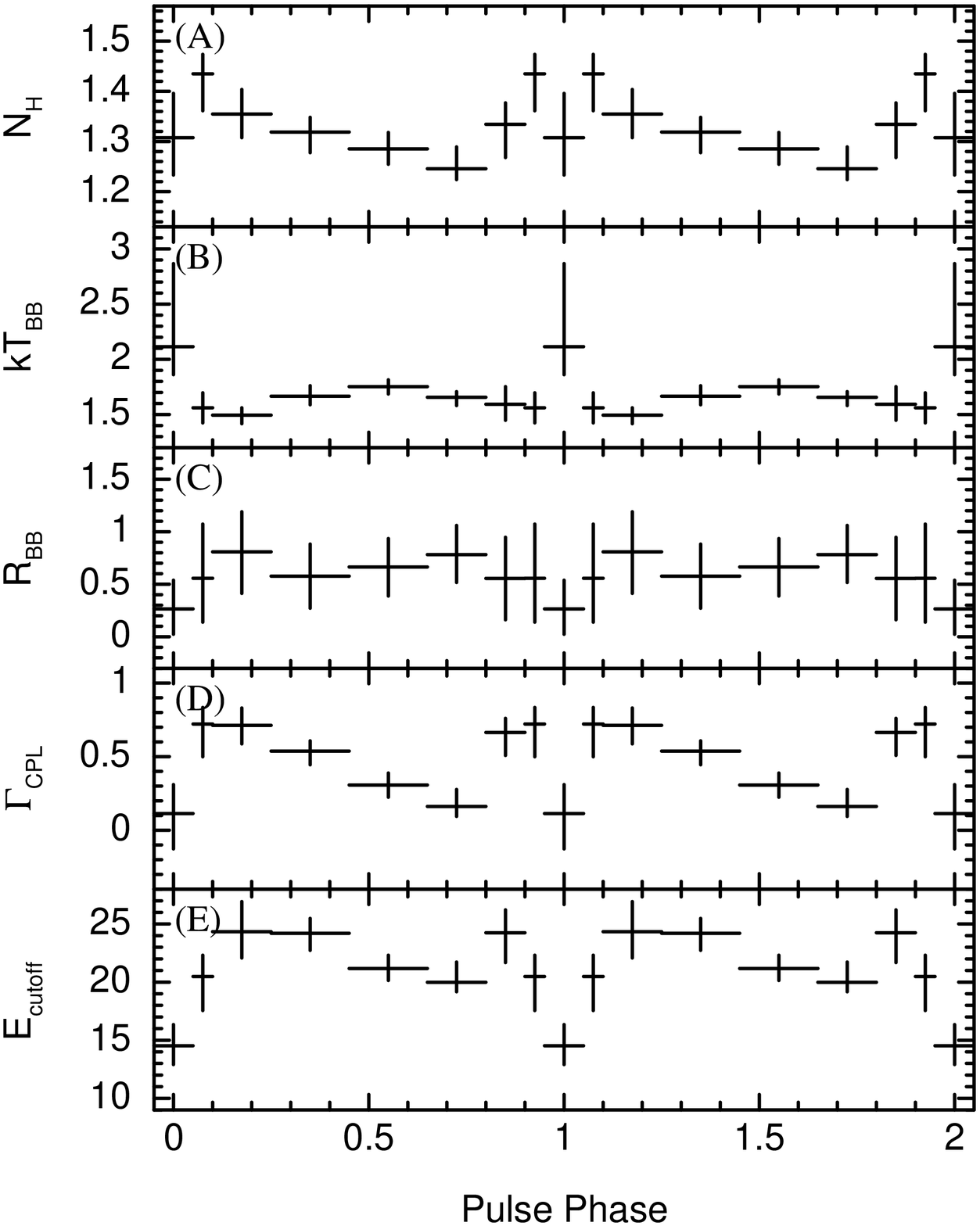}{0.3\textwidth}{}
	\fig{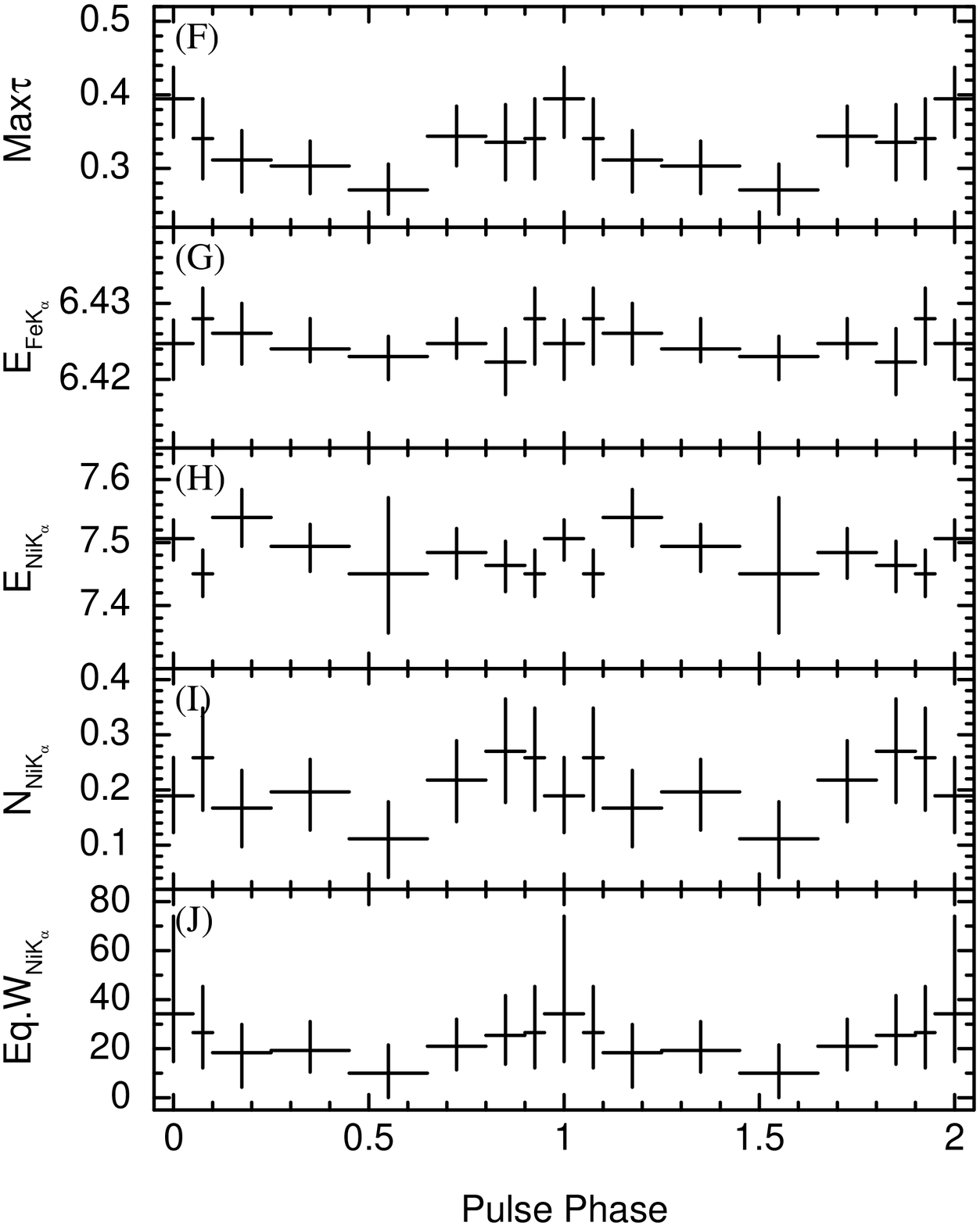}{0.3\textwidth}{}
	     \fig{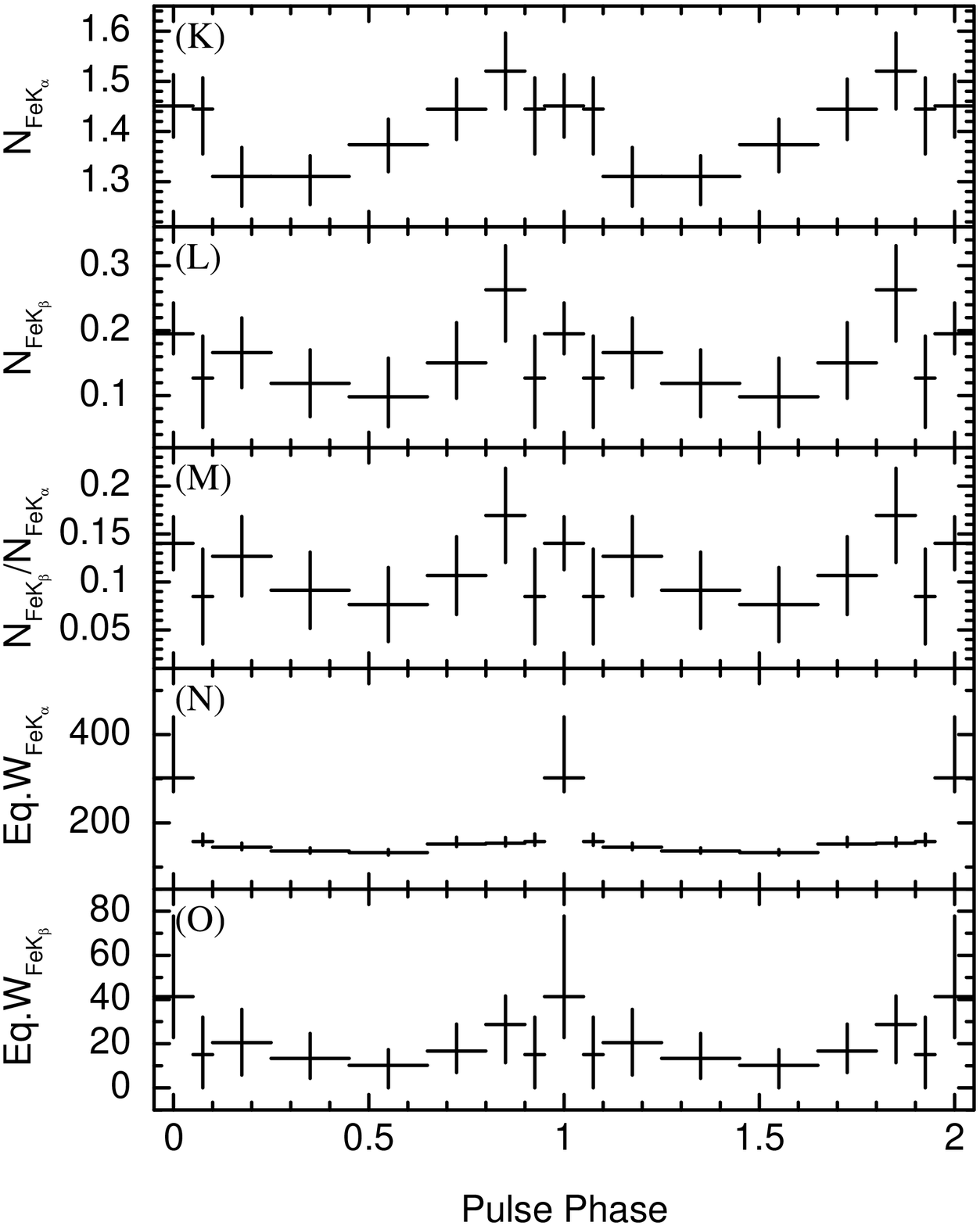}{0.3\textwidth}{}}
   \caption{
   Results of phase-sliced spectroscopy. 
   The spectral parameters were plotted as a function of pulse phase. 
   The parameters shown in each of the panels are described as follows. 
   \edit1{(A) photoelectric absorption column density (in H\,atoms\,10$^{23}$\,cm$^{-2}$),
   (B) blackbody temperature (in keV),
   (C) radius of blackbody (in km; assuming a distance of 4.3~kpc.),
   (D) photon index of exponential cutoff power-law component,
   (E) cutoff energy of exponential cutoff power-law component (in keV),}
   (F) maximum optical depth of the iron K edge,
   (G) center energies of the iron K$_\alpha$ line (in keV),
   (H) center energies of the Ni\,K$_\alpha$ line (in keV),
   (I) intensities of the Ni\,K$_\alpha$ line (in 10$^{-3}$\,photons\,s$^{-1}$\,cm$^{-2}$),
   (J) equivalent widths of the Ni\,K$_\alpha$ line (in eV),
   (K) intensities of the iron K$_\alpha$ line (in 10$^{-3}$\,photons\,s$^{-1}$\,cm$^{-2}$),
   (L) intensities of the iron K$_\beta$ line (in 10$^{-3}$\,photons\,s$^{-1}$\,cm$^{-2}$),
   (M) intensity ratios of the iron K$_\beta$ line to the iron K$_\alpha$ line,
   (N) equivalent widths of the iron K$_\alpha$ line (in eV) and
   (O) equivalent widths of the iron K$_\beta$ line (in eV).}
    \label{fig:PhaseDivParam}
\end{figure*}

We fitted the phase-sliced spectra in the 2--120 keV ranges with the model consisting of 
$\texttt{TBabs}\times (continuum +\texttt{gauss}_{\rm{Fe\,K}_{\alpha}}+\texttt{gauss}_{\rm{Fe\,K}_{\beta}}+\texttt{gauss}_{\rm{Ni\,K}_{\alpha}}$). As in case of phase-averaged spectroscopy, we used BB+CPL continuum model to fit 
all the phase-sliced spectra. It was found that above model fitted all seven spectra very well 
and the residuals obtained from our fitting are shown in the left panels of 
Figure~\ref{fig:gxFitResultPhaseDiv}. The reduced $\chi^2$ values obtained were in the range 
of 1.15 to 1.37. The best-fit parameters derived from phase-sliced spectral fitting by using 
BB+CPL continuum model are listed in Table~\ref{tab:suzaku-phase-sliced} and plotted as function 
of pulse phase in Figure~\ref{fig:PhaseDivParam}. The column density ($N_{\rm H}$) of photoelectric 
absorption was found to be constant (within errors) over pulse phases while the value of photon 
index and cutoff energy showed variations over pulse phase. One notable result of this fitting is 
that the $kT_{\rm BB}$ and $\Gamma$ at the dip interval are significantly larger and flatter than 
those at other phases as shown in Figure~\ref{fig:PhaseDivParam}(B) and (D).

As an alternative representations of the BB+CPL $continuum$, we adopted the \texttt{compTT} model 
to the phase-sliced spectra as was used to reproduce the source spectrum in previous works. 
The residuals obtained from the fitting by the \texttt{compTT} continuum model are shown in 
Figure~\ref{fig:gxFitResultPhaseDiv} and parameters obtained from spectral fitting are compiled 
in Table~\ref{tab:suzaku-phase-sliced}. The values of reduced $\chi^2$ obtained from each of the 
phase-sliced spectral fitting with the \texttt{compTT} model were $<2$. The equivalent hydrogen 
column density did not show any significant variation over pulse phases of the pulsar. While the 
optical depth of hot plasma $\tau$ and the photon source temperature $T_{\rm o}$ during the dip 
interval resulted in larger values than those of other phases, the hot electron temperature 
$T_{\rm e}$ and the normalization $A_{\rm{c}}$ were minimum at the dip interval. However, as in 
the case of phase-averaged spectroscopy, the residuals show wavy structures above 20~keV which 
yielded larger values of reduced $\chi^2$ than those obtained from the BB+CPL model.

To investigate the variation of emission lines and related features with pulse phase, we used same model as in case of phase-averaged spectroscopy e.g. power-law continuum multiplied by an absorption edge and the emission lines (iron K$_{\alpha}$, iron K$_{\beta}$ and Ni K$_{\alpha}$ lines) multiplied by the photoelectric absorption (\texttt{TBabs}) to fit the phase-sliced spectra. 
Then, we fixed the ratios of the energies of the iron K$_{\beta}$ line and the edge to that of the iron K$_{\alpha}$ line at 1.103$\times$ and 1.119$\times$, respectively and the widths of three lines were fixed to be narrow enough according to the results of the phase-averaged spectroscopy. 
The values of equivalent hydrogen column density $N_{\rm H}$ were fixed at (1.29 -- 1.38) $\times10^{23}$~H\,atoms\,cm$^{-2}$ for each phase-sliced spectrum (see 
Table~\ref{tab:suzaku-phase-sliced}). 
The iron Ly$_\alpha$ line was not included in this analysis.

Parameters obtained from our spectral fitting are plotted in Figure~\ref{fig:PhaseDivParam} and summarized in Table~\ref{tab:suzaku-phase-sliced}. 
The center energies of the iron ($E_{{\rm Fe\,K}_{\alpha}}$) and the Ni lines ($E_{{\rm Ni\,K}_{\alpha}}$) were found to remain constant. 
Enhancement of the equivalent width of iron K$_{\alpha}$ line at the dip interval was notable and the same was seen for iron K$_{\beta}$ and Ni lines. 
The intensity ratio of the iron K$_{\beta}$ to the K$_{\alpha}$ lines were consistent with a constant value (within errors) at 0.12 of the neutral case \citep{Yamaguchi2014}. 
An intensity modulation of the iron K$_{\alpha}$ line can be seen in the figure. 
The line became intense in the phase between 0.7 and 1.1. Intensities of iron K$_{\beta}$ and the Ni K$_{\alpha}$ lines also showed similar behavior. 
\edit1{We performed $\chi^{2}$ test to investigate the presence/absence of modulation in the line intensity with pulse phase, which yielded $\chi^2 / d.o.f. = 29.84 / 7$. 
This value indicated that the modulation of line intensity with pulse phase is statistically significant with a null hypothesis probability of 1.0$\times 10^{-4}$. }
The depth of the edge was slightly shallow at around phase 0.5 although the errors are large.

\section{DISCUSSION} \label{sec:discuss}

\subsection{Pulse Profile and Dip} \label{subsec:dip}

Pulse profiles of GX~1+4 have been studied by several authors to understand the 
cause of the presence of peculiar sharp dip. Detailed investigation of the source
properties during the dip can provide important information about accretion flow 
and geometry of the emission region. Dips in pulse profiles can be originated by 
various mechanisms which were discussed in previous studies. \citet{Dotani1989} 
interpreted that the dip structure in the pulse profile is due to the cyclotron 
resonant scattering of photons at the accretion column above the magnetic pole. 
Recently it has been reported that the cyclotron resonance scattering can affect 
the pulse profile of accretion powered X-ray pulsars at energies closer to the 
cyclotron absorption line energy, as seen in case of Be/X-ray binary pulsar 
GX~304-1 (\citealt{Jaisawal2016} and references therein). However, cyclotron 
absorption line has not yet been detected in the broad-band spectra of GX~1+4. 
Therefore, the cyclotron resonance scattering feature may not be the cause of 
sharp dip seen in the pulse profile of GX~1+4. In phase-resolved spectroscopy 
of the {\it RXTE} observations, \citet{Galloway2001} found an increase in optical 
depth at dip phase of the pulsar. They interpreted the sharp and prominent dip in 
the pulse profile as due to the obscuration or eclipse of hot spot by the 
accretion column. Similar explanation was also proposed by \citet{Giles2000} 
during a faint state of GX~1+4. 

We carried out phase-averaged and phase-sliced spectroscopy to investigate the 
changes in spectral parameters during the dip and non-dip phases of the pulsar 
by using high quality data from {\it Suzaku} observation. Spectral hardening was 
observed during the dip interval compared to the non-dip phases. However, the 
value of equivalent hydrogen column density did not show any significant variation 
within 90\% confidence level. Moreover, the spectral ratio for the dip interval 
(Figure~\ref{fig:gxPhDivSpecHR}) showed the dimming of the source over wide energy 
band, especially below 10~keV. These results indicate that the dip in the pulse 
profile of GX~1+4 is not due to the increase in the photoelectric absorption, rather 
due to the scattering of photons by hot electrons in optically thick region compared 
to other phase intervals. This supports the earlier interpretation of alignment of 
the accretion column containing hot plasma with the line of sight as the cause of 
sharp dip in the pulse profile of GX~1+4.

\citet{Naik2005} pointed out the widening of the dip at higher energies and speculated 
that the high-energy photons escape from the column preferentially at large angle whereas 
the low-energy photons are more isotropic. In other words, the emission geometry might 
have changed from a pencil beam at low energy to a fan beam at high energy. The 
energy-resolved pulse profiles of GX~1+4 obtained from {\it Suzaku} observation 
also indicated similar shape change with energy. The idea by \citet{Naik2005}, 
therefore, can qualitatively explain the observed behavior. A similar idea had 
been proposed by \citet{GallowayWu1999} where photons emitted from the polar caps 
get Compton scattered by the plasma in the accretion column before being escaped 
towards the observer.

\subsection{Broad-band Spectroscopy of GX~1+4}\label{sec:EnergySpectra}
\subsubsection{X-Ray Spectrum with Suzaku Observation}

In intermediate and high luminosity states, the spectra of GX~1+4 have been 
described by the \texttt{compTT} model \citep{Galloway2000,Naik2005,Ferrigno2007}. 
The luminosity during {\it Suzaku} observation of the pulsar was comparable to 
those observations. Considering earlier results, we also attempted \texttt{compTT} 
continuum model to fit the observed phase-averaged broad-band spectrum. The results 
obtained from our spectral fitting were found to be consistent with those reported 
earlier. However, presence of wavy structures in the residuals obtained from fitting 
the source spectra with \texttt{compTT} continuum model allowed us to try another 
empirical model e.g. blackbody and cutoff power-law (BB+CPL) model in spectral 
fitting. This model improved the spectral fitting without showing any such features 
in the residuals. 

The \texttt{compTT} continuum model provides meaningful physical parameters compared 
to other empirical models. In our phase-sliced spectroscopy with \texttt{compTT} 
continuum model, the optical depth $\tau$, the photon source temperature $T_{\rm o}$, 
the hot electron temperature $T_{\rm e}$ and the normalization $A_{\rm{c}}$ were found 
to be significantly modulated with pulse phase of the pulsar. In particular, the 
increasing value of $\tau$ and decreasing values of $T_{\rm e}$ and $A_{\rm{c}}$ 
at the dip interval were also reported by \citet{Galloway2000}.

In the \texttt{compTT} model, the seed photons undergo scattering by thermal hot 
plasma through an escape probability distribution which depends on optical depth 
and geometry (either sphere or disk). Therefore, this model is insufficient to 
describe Comptonization in accretion column where the hot plasma has a cylindrical 
geometry with base as the source of seed photons. Additional parameters should be 
introduced, which affect the energy spectra such as the ratio of height and radius 
of the cylinder and the angle between the line of sight and the cylinder axis.
For example, only scattered photons are expected to be observed from the column 
if it is viewed at right angle with respect to cylinder axis, even if the optical 
depth is $\le$1. On the other hand, the number of scattered photons gets reduced 
when viewed along the cylinder axis. Therefore, viewing angle of the accretion 
column plays a crucial role in shaping the energy continuum. Better fitting of 
{\it Suzaku} data with BB+CPL continuum model suggested that cylindrical geometry 
is most preferred in GX~1+4. In this geometry, the observed spectrum can be 
separated into two components e.g. seed photon component described by blackbody 
(BB) and the scattered photon component expressed as cutoff power-law (CPL). 
We also tried to fit the phase-averaged broad-band spectra of GX~1+4 with two 
component models such as \texttt{compTT}+ \texttt{compTT} (used to describe hard 
X-ray spectrum of Be/X-ray binary pulsar X~Per by \citet{Doroshenko2012}) and 
\texttt{compTT}+BB. These models also fitted the energy spectra well with 
comparable values of reduced $\chi^{2}$ as in case of BB+CPL model. Although 
these empirical models fit the data better, the correspondence between the 
obtained parameters and actual physical ones is not clear. For this purpose, 
more sophisticated modeling is required by considering the scattering cross 
section in strong magnetic field and realistic geometry of the accretion column.

\subsubsection{\edit1{X-Ray Spectra of Previous Observations}}
\edit1{In an effort to confirm our interpretation, we analyzed archival data from the {\it Beppo}SAX and Rossi X-ray Timing Explorer ({\it RXTE}) observations of the pulsar and then compare the results obtained from the {\it Suzaku} observation of GX~1+4. 
The log of these observations is given in Table~\ref{tab:xslog}. 
We have used two {\it Beppo}SAX observations of the pulsar GX~1+4 when the source was at a flux level of $\sim$10$^{-9}$ erg\,s$^{-1}$\,cm$^{-2}$ in 2-100 keV energy range. 
These two observations are the same as used by \citet{Naik2005}. 
The data were obtained with three major instruments such as Low Energy Concentrator Spectrometer (LECS; 0.1--5 keV), 
Medium Energy Concentrator Spectrometers (MECS; 1--10 keV), and hard X-ray Phoswich Detector System (PDS; 15--300 keV), covering a broad energy range from soft to hard X-rays \citep{Boella1997}. 
During the observation in 1996, LECS was not operated and the effective exposures for MECS and PDS were 38.6 and 17.6~ks, respectively. 
The 1997 observation was carried out for effective exposures of 13.0, 31.5 and 13.5~ks for LECS, MECS, and PDS, respectively. 
Standard procedures were followed for data reduction for both the {\it Beppo}SAX observations of GX~1+4. 
Spectra from LECS and MECS were extracted from CCD chips by selecting circular regions of 4' around the source center. Background spectra for these observations were also extracted by selecting circular regions away from the source. 
The PDS spectra were retrieved from the standard products of both the observations. 
Using appropriate spectra, background, and response files as provided by instruments teams, spectral fitting was carried out in $\sim$1--150 keV range.}

We also analyzed archival data of the pulsar obtained from the Proportional Counter Array (PCA; \citealt{Jahoda1996}) and High Energy X-ray Timing Experiment (HEXTE; \citealt{Rothschild1998}) onboard the {\it RXTE} satellite to obtain a suitable continuum model.
For this purpose, four {\it RXTE} observations of the pulsar with long exposures ($>$15~ks), during intermediate and high luminosity phases were chosen to examine the properties of the pulsar. 
One of these observations (X-I\hspace{-.1em}I\hspace{-.1em}I) is the same {\it RXTE} observation, the results of which are published in \citet{Galloway2001}. 
HEAsoft analysis package (version 6.16) and up to date calibration data base (CALDB) files were used during reduction of {\it RXTE} data. 
For our study, we extracted source and background spectra from Standard-2 mode PCA data following standard procedure. Data from all available PCUs were used in our analysis. 
Using data from HEXTE Cluster-A, source and background spectra were obtained by using standard tasks of FTOOLS. The dead time correction was also applied to the HEXTE spectra. 
The response files for PCA and HEXTE detectors were created by using \texttt{pcarsp} and \texttt{hxdrsp} commands, respectively. In the spectral fitting, data in the range of 3--150 keV were used from the RXTE observations.

\begin{figure}
\begin{center}
   \includegraphics[height=60mm]{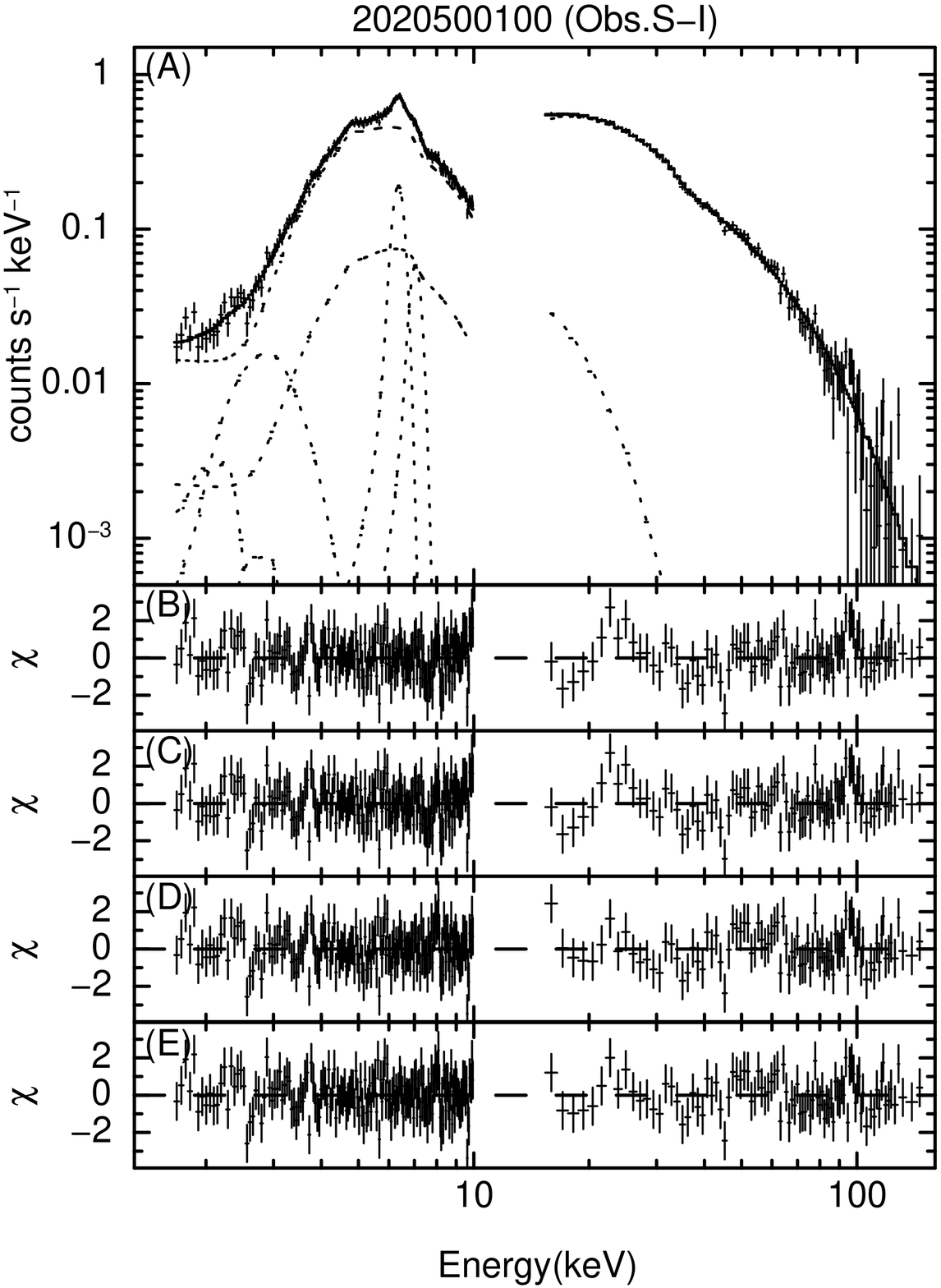} \hspace{-.7em}
   \includegraphics[height=60mm]{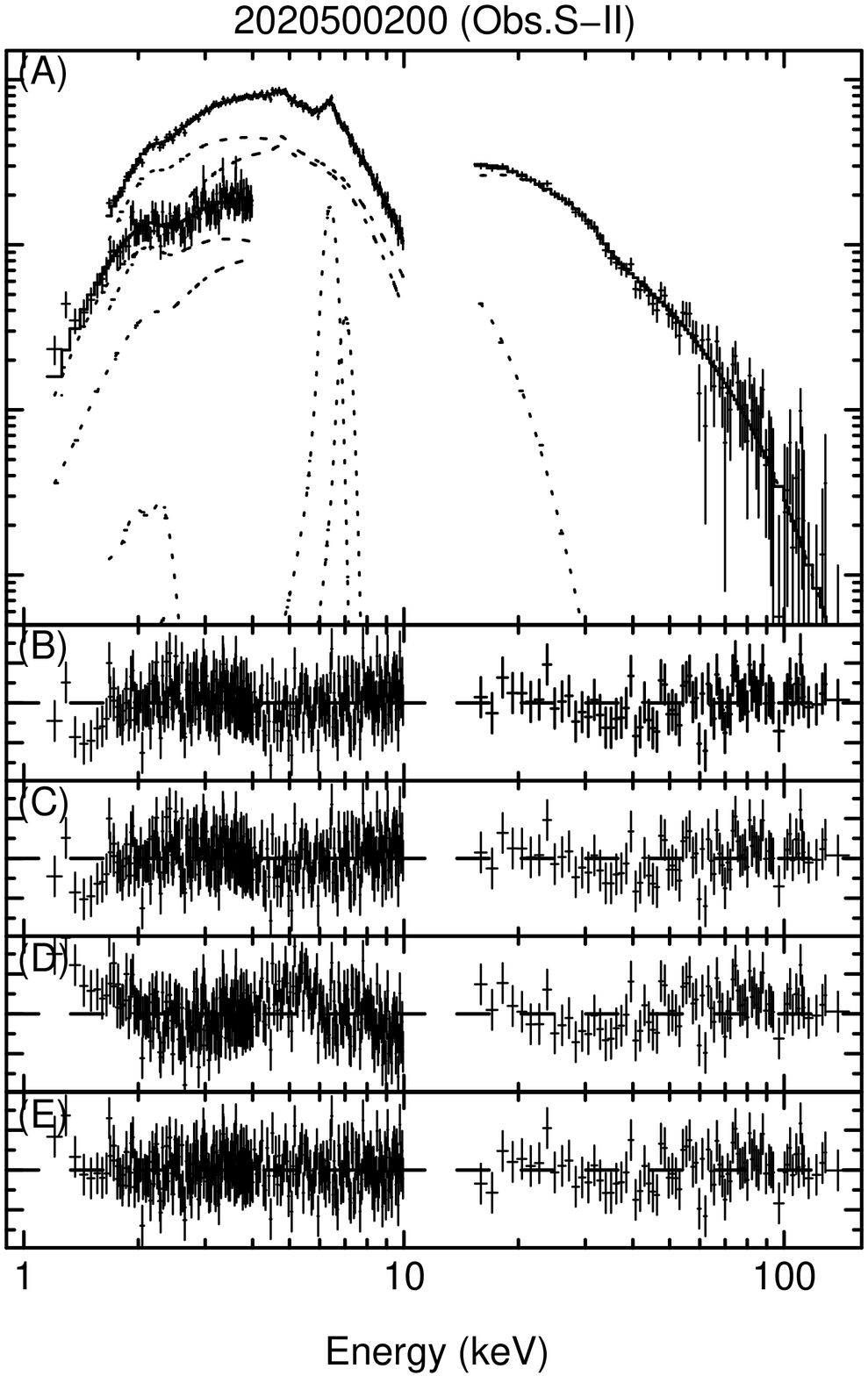} \\
   \vspace{.7em}
   \includegraphics[height=60mm]{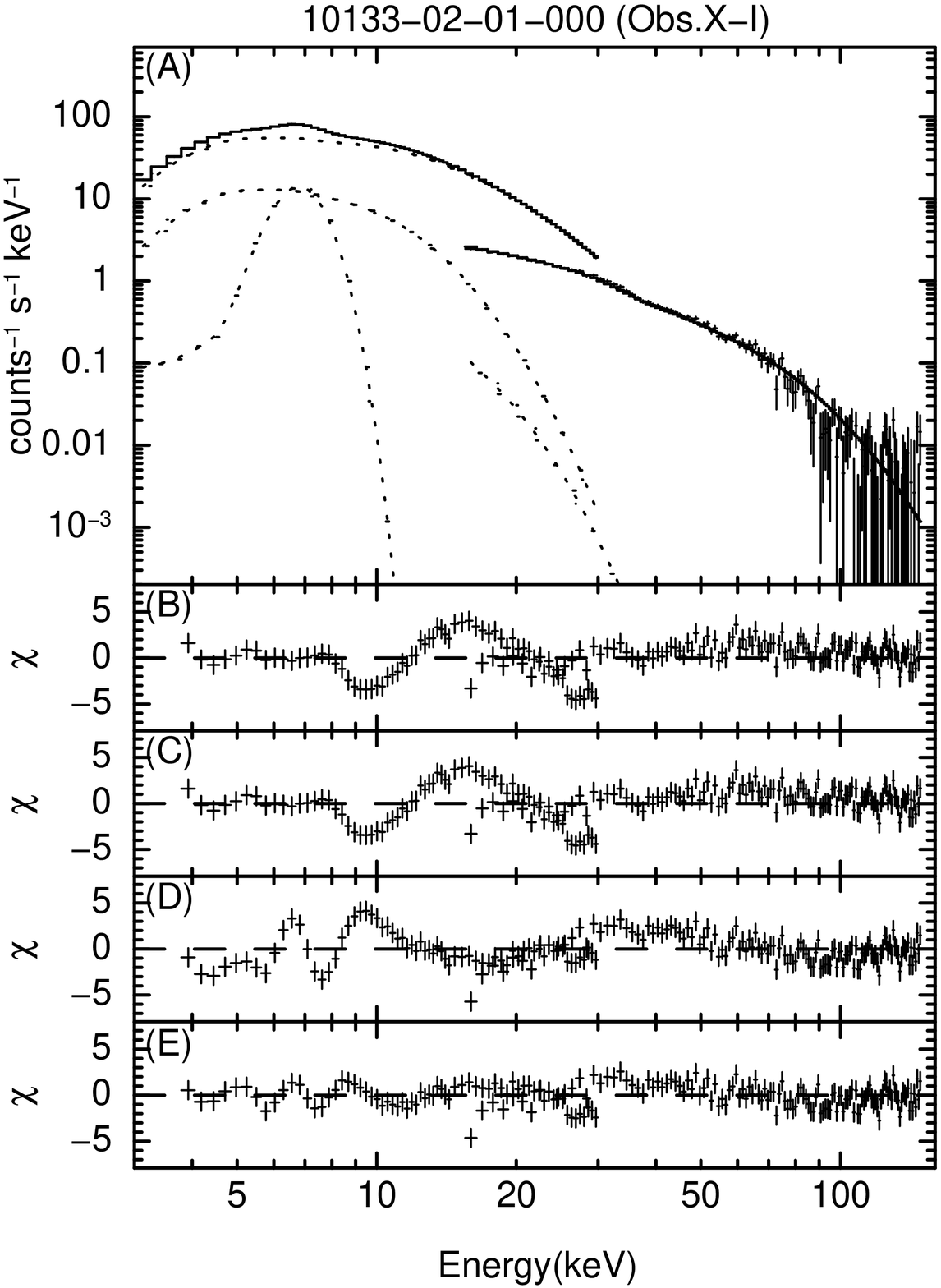} \hspace{-.7em}
   \includegraphics[height=60mm]{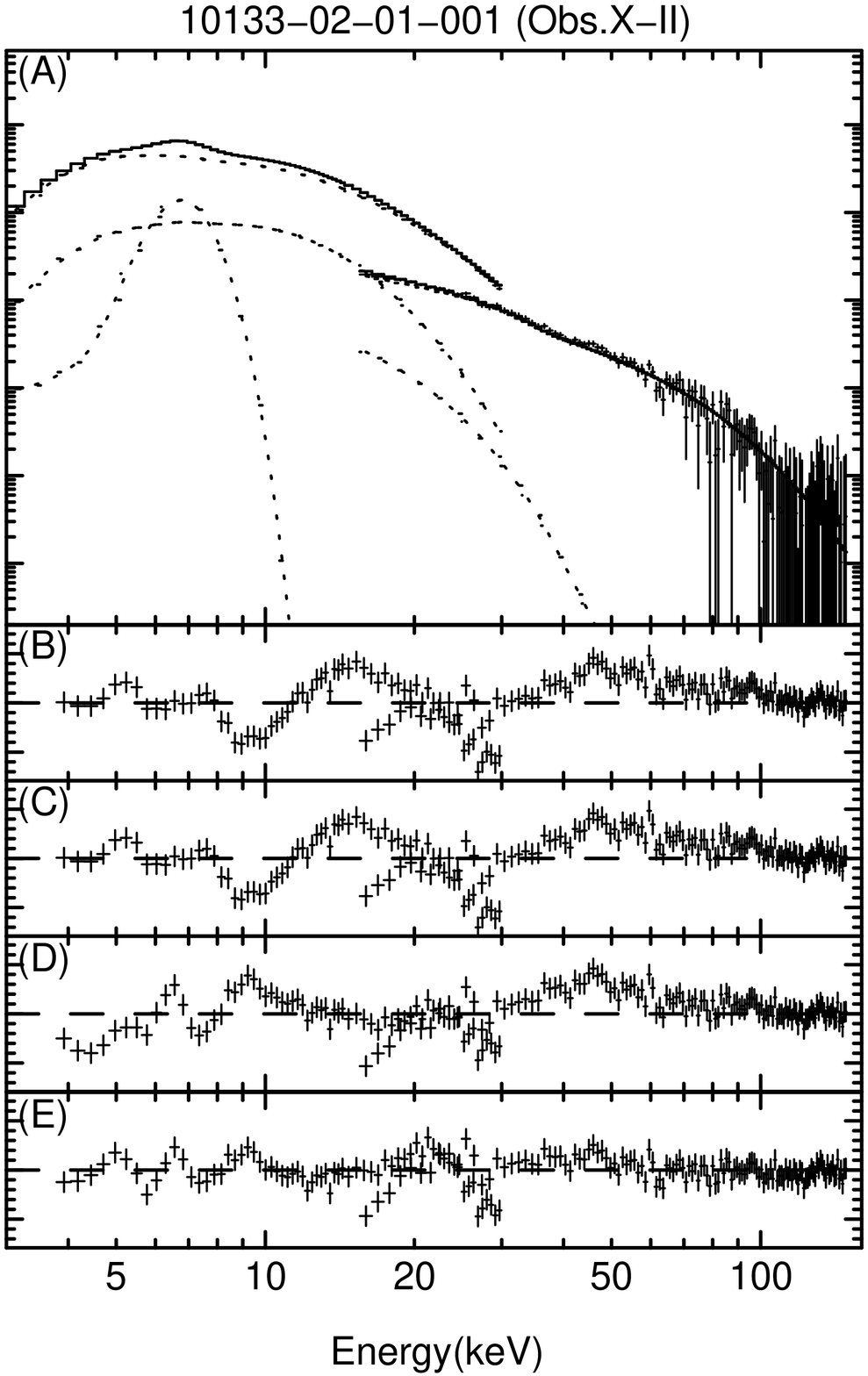} \\
   \vspace{.7em}
   \includegraphics[height=60mm]{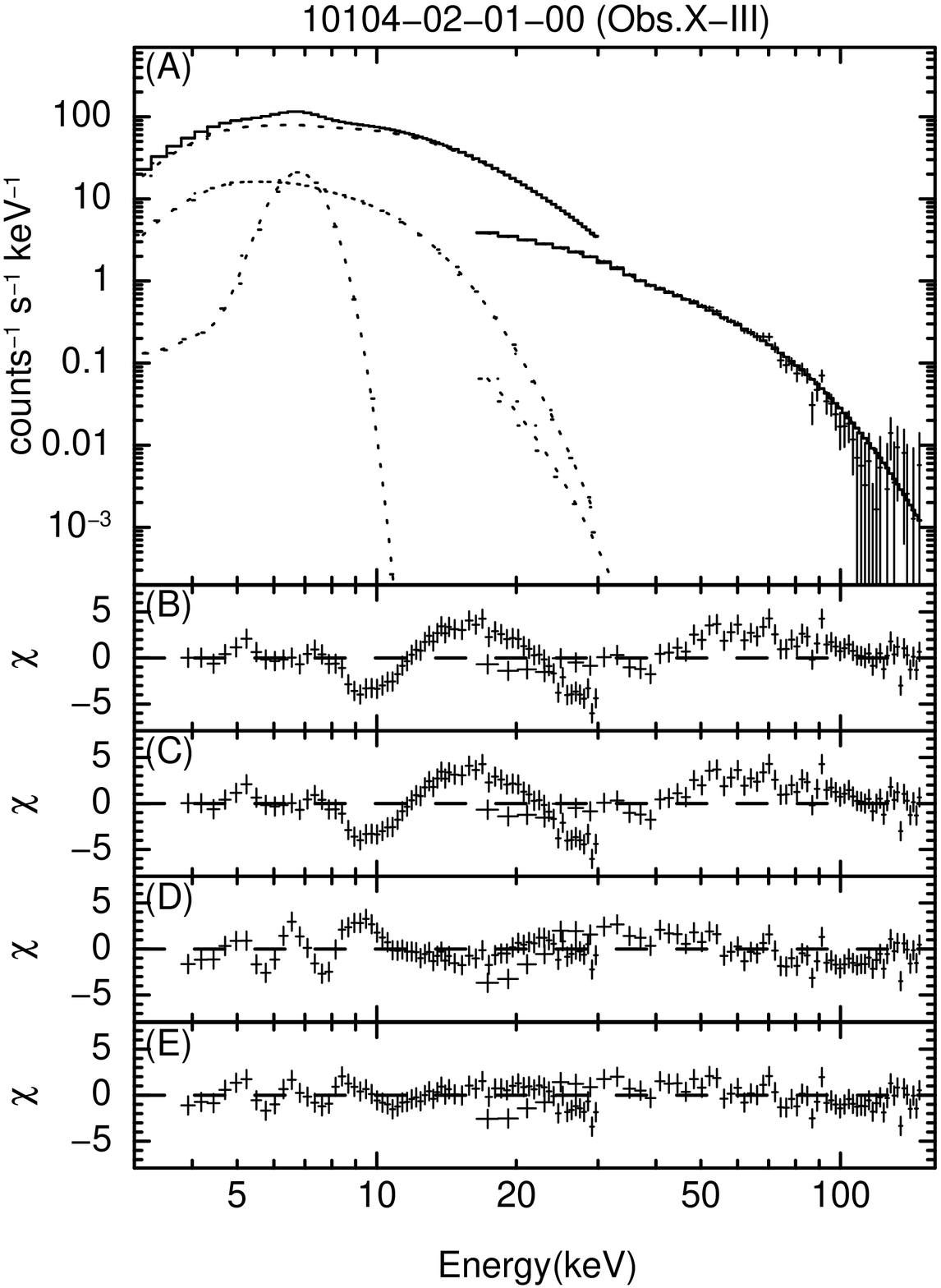} \hspace{-.7em}
   \includegraphics[height=60mm]{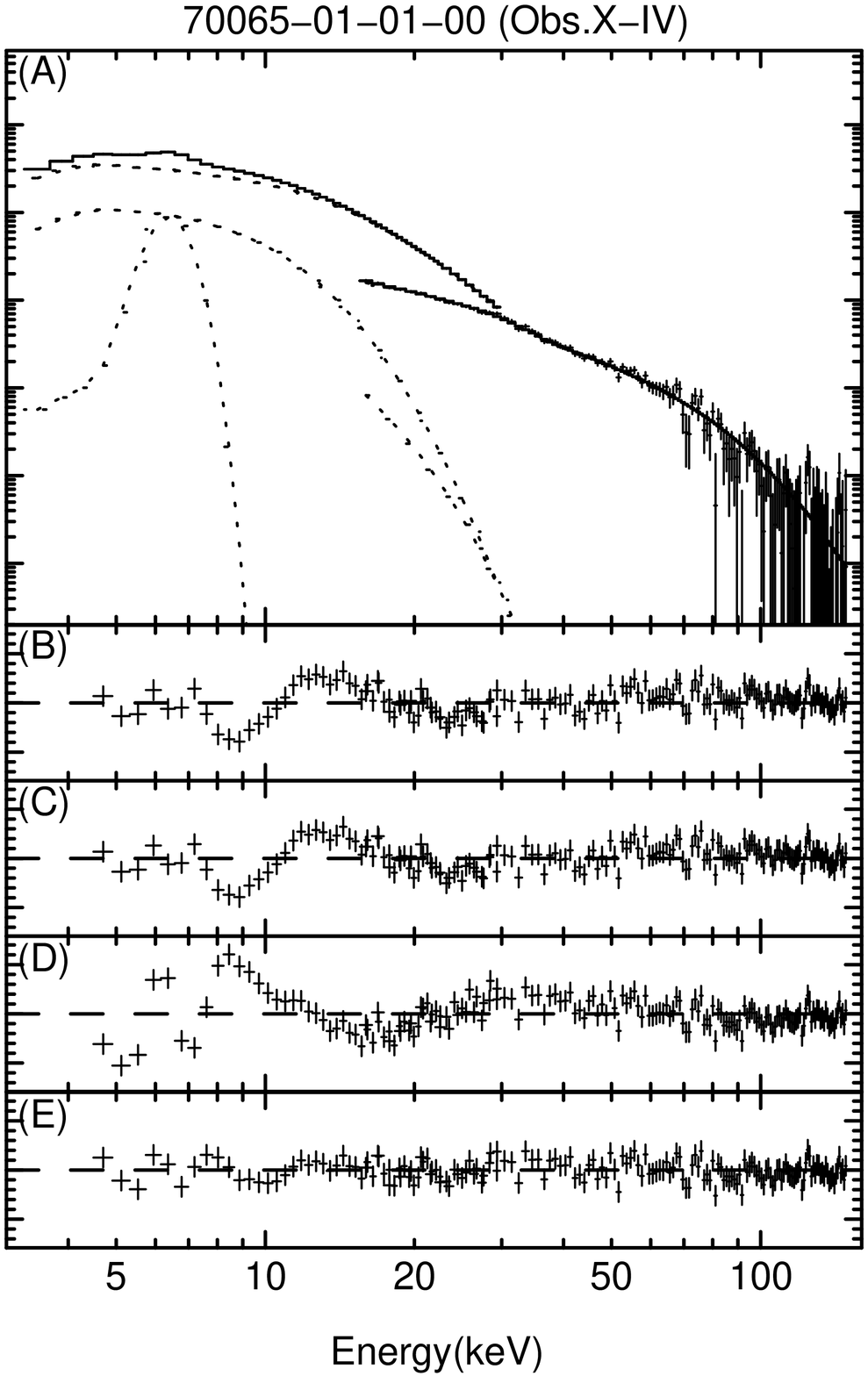}
   \end{center}   
   \caption{
   Broad-band phase-averaged spectra of GX~1+4 obtained from the {\it Beppo}SAX and the {\it RXTE} observations along with the best fit BB+CPL model are shown in top panels (A). 
   \edit1{Residuals divided by its errors}, obtained from fitting the source spectra with \texttt{compTT} model assuming disk and spherical geometry are shown in panels (B) and (C) whereas those obtained from fitting data with CPL and BB+CPL models, are shown in panels (D) and (E), respectively. 
   Each of the six figures show the results, which were obtained from fitting with different continuum models for six epochs (two epochs of {\it Beppo}SAX and four epochs of {\it RXTE}) observations in the sequence of S-I,  S-I\hspace{-.1em}I, X-I, X-I\hspace{-.1em}I, X-I\hspace{-.1em}I\hspace{-.1em}I and X-I\hspace{-.1em}V as given in Table~\ref{tab:xslog}. The sequences are quoted above each individual figure.}
   \label{fig:gxXSFitResultPhaseAve}
\end{figure}

We fitted all the phase-averaged broad-band spectra obtained from the {\it Beppo}SAX and {\it RXTE} observations (Table~\ref{tab:xslog}) by using four continuum models such as \texttt{compTT} with a geometry of a disk, \texttt{compTT} with a geometry of a sphere, CPL, and BB+CPL. In all the models, 
a component for photoelectric absorption (\texttt{TBabs}) and Gaussian functions for iron emission lines were included. 
We added 0.5~\% systematic errors to the {\it RXTE} spectra. 
\edit1{In {\it Beppo}SAX data analyses, we did not include the edge component at 34~keV which was reported by \citet{Naik2005}. 
While the fitting data from the 1996 {\it Beppo}SAX observation, we added a bremsstrahlung component to the spectral model representing the soft X-ray excess as reported by 
\citet{Naik2005}. }
In the simultaneous spectral fitting, we found that the BB+CPL model described the continuum spectra of the pulsar obtained from both the observatories better than other models, as in case of {\it Suzaku} data. 
The broad-band spectra along with the best-fit BB+CPL model for six epochs of observations are shown in the top panels of Figure~\ref{fig:gxXSFitResultPhaseAve}. 
Residuals of the data from the model obtained from fitting the phase-averaged spectra for six epochs with \texttt{compTT} continuum model for a disk geometry, \texttt{compTT} for a spherical geometry, cutoff power-law continuum model and a blackbody and cutoff power-law model, are shown in panels (B), (C), (D), and 
(E) of Figure~\ref{fig:gxXSFitResultPhaseAve}, respectively. 
Best-fit parameters obtained for the \texttt{compTT} model (for disk and sphere geometries) and the BB+CPL model are given in Table~\ref{tab:xslog}.

\subsection{Origin and Region of Line Emission}
\subsubsection{Line Center Energy}
Using {\it Suzaku} observation, we detected iron K$_{\alpha}$, K$_{\beta}$ emission lines and K absorption edge in GX~1+4. 
The ratio of the energy of the K absorption edge to that of the K$_{\alpha}$ emission line strongly restricts the ionization state of the iron ion to be less than 2 ($<$Fe\,I\hspace{-.1em}I\hspace{-.1em}I). 
However, the absolute energy of iron K$_{\alpha}$ emission line e.g. 6.425$\pm$0.001~keV, as observed in present study, is not consistent with the energy of neutral iron in a laboratory frame and is higher by approximately $+$30 eV. 
Similar values of line energy were also reported by \citet{Naik2005} from {\it Beppo}SAX observations of the pulsar. 
However, using {\it Chandra}/HETGS observation of GX~1+4, \citet{Paul2005} reported 6.400$\pm$0.005~keV as the energy corresponding to Fe K$_{\alpha}$ emission line. 
Although we applied the correction for SCF effect, the absolute energy determination with a CCD is extremely difficult. 
The observed discrepancy in the line energy of iron K$_{\alpha}$ emission line is comparable to the energy calibration ambiguity. 
Therefore, future observations with high spectral capability instruments are required to understand this energy shift.

\subsubsection{Homogeneous Matter} 
The emission region of the fluorescent iron line has been discussed by \citet{Kotani1999}, 
where the matter is assumed to be homogeneously distributed. They showed a positive 
correlation between the iron emission line equivalent width and absorption column density 
which was consistent with the expected results from isotropic distribution of absorbing 
matter around the pulsar \citep{Makishima1986}. The values of equivalent width ($\sim$150~eV) 
and absorption column density ($1.30\times10^{23}$~H\,atoms\,cm$^{-2}$) obtained from 
{\it Suzaku} observation of the pulsar are also consistent with the relation. During the 
{\it Suzaku} observation, the ionization state of iron atoms was determined to be 
$\le$Fe\,I\hspace{-.1em}I\hspace{-.1em}I. This corresponds to the value of ionization 
parameter $\xi(\equiv\frac{L}{nr^2})$ to be less than 22.4 ($\log \xi <1.35$) \citep{Kallman1982} 
for an ionizing source of 10$^{37}$ erg\,s$^{-1}$ with a 10~keV bremsstrahlung in optically 
thick plasma. Luminosity of GX~1+4 during the {\it Suzaku} observation was estimated to be 
$10^{37}$~erg\,s$^{-1}$ (assuming a distance from the source of 4.3~kpc) and the value of 
equivalent hydrogen column density was $1.30\times10^{23}$~H\,atoms\,cm$^{-2}$. Using these 
values and assuming that the distance from the X-ray source is comparable to the size of the 
plasma, the location of fluorescent line emitting region from the X-ray source is estimated 
to be more than 3.4$\times$10$^{12}$~cm, which is consistent with the value reported by 
\citet{Kotani1999}. This value is comparable to the size of the orbit ($\sim10^{13}$ cm) for 
the earlier reported period of 300~d \citep{Cutler1986,Pereira1999}. If the fluorescent lines 
are emitted from such a large region, the observed time variation should be smeared with the 
light crossing time of the region, which is more than 100~s for the size of 3.4$\times$10$^{12}$~cm. 
One of our new finding is the intensity modulation of the fluorescent line with the neutron 
star rotation. The line was intense in the pulse phase range of 0.7--1.1 with an amplitude of 7\%. 
As the modulation amplitude is not large and the spin period of pulsar is $\sim$150~s, observed 
intensity modulation might be possible as long as the plasma size is not significantly larger 
than 3.4$\times$10$^{12}$~cm. If it is caused by a non-symmetric circumstellar matter with the 
size of the binary orbit, we can expect a progressing of the line-intense-phase according to 
the orbital motion and we may examine this idea by using future observation.

\edit1{There are only a few sources which show pulse phase dependence of emission line flux viz. flux of O\,V\hspace{-.1em}I\hspace{-.1em}I line in 4U~1626-67 \citep{Beri2015} and fluorescent iron emission line in Her~X-1 \citep{Vasco2013} and GX~301-2 \citep{Suchy2012}.
In the case of 4U~1626-67 ($P_{spin}\sim$7.7~s), the intensity modulation by a factor of about 4 over pulse phases was reported \citep{Beri2015}. 
This was interpreted in terms of a warped accretion disk being illuminated by the X-rays from the neutron star. 
In case of Her~X-1 ($P_{spin}\sim$1.24~s), the fluorescent iron line intensity shows a sharp and deep minimum (close to zero) at the peak of the pulse profile \citep{Vasco2013}. 
Based on these results, \citet{Vasco2013} discussed the possibility of accretion column being the line emitting region in Her~X-1. 
For these fast rotators, the emission regions are considered to exist at the vicinity of the neutron star. 
However, GX~301-2 is a slow rotator with a spin period of $P_{spin}\sim$700~s and rather sinusoidal intensity modulation of iron fluorescent line was reported with a modulation amplitude of ~10\% \citep{Suchy2012}. 
In this case, the line emitting region is considered to be very large. 
In GX~1+4, we found sinusoidal modulation of the iron line intensity at an amplitude of $\sim$7\%. Maximum of the line intensity does not coincide with either the minimum or maximum of the continuum emission over pulse phases.
These characteristics are similar to those seen in GX~301-2.}

\subsubsection{Inhomogeneous Matter}
Size of the line emitting region, as discussed in previous subsection, was derived by assuming homogeneous distribution of matter and the observed ionization state. 
However, if we introduce a volume filling factor $f$, the result becomes completely different, \edit1{though this is just one possibility}. 
Assuming $n$ as the density of highly 
dense region and negligible (close to zero) density for a thin region, the observed 
column density can be written as $N_{\rm H}\sim nfr$, where $r$ is the size of fluorescent 
region. This $r$ can also be assumed as the distance of the region. The ionization 
parameter of the dense region, $\xi$, can be expressed as $\frac{L}{nr^2}$. For 
$\xi=22.4, L=10^{37}$ erg\,s$^{-1}$, and $N_H=1.30\times 10^{23}$~H\,atoms\,cm$^{-2}$, 
we obtained the distance $r>(3.3 f)\times 10^{12}$ cm. For $f \sim 0.03$, the size of 
the fluorescent region can be reduced to $\sim10^{11}$~cm which is comparable to the 
size of an accretion disk.

\citet{Paul2005} reported the detection of Ly$_{\alpha}$ emission line from iron 
ions in GX~1+4 and suggested that the diffused gas in the Alfv\'en sphere or the 
accretion curtains to the magnetic poles as the line emitting regions. During the 
{\it Suzaku} observation of the pulsar, we also detected Ly$_{\alpha}$ emission 
line in the pulsar spectrum with comparable intensity as reported by \citet{Paul2005}. 
If we consider the inhomogeneous matter as the fluorescent line emitting gas, then 
the thin region should be highly ionized and may be the possible origin of the 
Ly$_\alpha$ emission line. Another possibility for the smaller size of the 
fluorescent region can be due to the presence of local dense matter. 
\citet{Rea2005} discussed the prospect of the existence of a thick torus-like accretion disk. 
If we introduce such a thick region surrounding the magneto-sphere with a radius 
of $\sim$8.2$\times$10$^9$~cm as a fluorescent region (assuming that the magnetospheric 
radius is equal to the co-rotation radius based on torque reversal scenario), the 
density of the torus should be $6.5\times10^{15}$~cm$^{-3}$ for $\xi=22.4$.

\subsection{Elemental Abundance}

For the first time, we detected an emission line at 7.49$\pm$0.02 keV in the 
spectrum of GX~1+4. This line was identified as K$_\alpha$ emission line from 
neutral Ni atoms which is expected at 7.47~keV in the laboratory frame. As this 
line is expected from neutral atoms, we can assume that Ni K$_\alpha$ emission 
also originated from the same fluorescent matter which is attributed to the iron 
K$_\alpha$ emissions. The abundance ratio [Ni/Fe] of the fluorescent matter can 
also be investigated by studying the properties of these lines. By assuming a 
small optical depth for the absorption, the expression for the ratio of line 
intensities can be written as, 

\begin{equation}
\frac{I_{\rm{Fe_{line}}}}{I_{\rm{Ni_{line}}}} \sim \frac{\eta_{\rm Fe} \int_{E_{\rm Fe_{th}}}^{\infty}I(E)\sigma_{\rm Fe}(E)N_{\rm H}A_{\rm Fe}\,{\rm d}E}
  {\eta_{\rm Ni} \int_{E_{\rm Ni_{th}}}^{\infty}I(E)\sigma_{\rm Ni}(E)N_{\rm H}A_{\rm Ni}\,{\rm d}E} ,
     \label{eq:intensityratio1}
\end{equation}

\noindent
where $I_{\rm{M_{line}}}$, $\sigma_{M}(E)$, $\eta_{\rm{M}}$, $A_{\rm{M}}$ and $E_{\rm{M_{th}}}$ 
are the intensity, photoelectric absorption cross section, fluorescence yield, elemental 
abundance and binding energy of the K-shell electrons for the neutral atom $M$. $I(E)$ 
is the continuum emission which is a function of the X-ray energy $E$ and $N_{\rm H}$ is 
the hydrogen column density. By putting appropriate parameters, Equation~\ref{eq:intensityratio1} can be reduced to,

\begin{equation}
\frac{I_{\rm{Ni_{line}}}}{I_{\rm{Fe_{line}}}} \sim 1.18 \frac{A_{\rm Ni}} {A_{\rm Fe}}.
     \label{eq:intensityratio2}
\end{equation}

\noindent
By inserting the observed value of intensity ratio of the emission lines of 0.12$\pm$0.02, 
we obtained $\frac{A_{\rm Ni}} {A_{\rm Fe}}$, i.e., [Ni/Fe] $=0.10\pm0.02$, which is 
approximately two times larger than that of the solar abundance \citep{Anders1989}.
Iron abundance can also be estimated by considering the depth of iron K-shell edge 
at around 7.1 keV. The maximum optical depth of the absorption edge was determined 
to be 0.31$\pm$0.02. If we assume the solar abundance of iron, $3.4\times10^{-5}$ 
and the absorption cross section of 9.2$\times10^{-20}$ cm$^2$, the expected optical 
depth is 0.4. As the maximum optical depth was 0.31$\pm$0.02, the iron abundance of 
the fluorescent matter is $\sim$80~\% of the solar value.

\section{Conclusion}

We carried out a detailed spectroscopic study of the X-ray emission from 
GX~1+4 by using {\it Suzaku} observation of the pulsar. Results obtained 
from the timing analysis, broad-band phase-averaged and phase-sliced spectroscopy, 
detailed investigation of the fluorescence emission lines and line emitting regions 
are summarized as follows:

\begin{enumerate}

\item 
Using {\it Suzaku} observation, the spin period of the pulsar was estimated to be 
$\sim$159.94~s. This indicates that the pulsar is spinning down. In addition, a 
peculiar sharp and prominent dip was also seen in the soft X-ray pulse profiles. 

\item 
The continuum spectrum of the pulsar can be described better by a two-component model consisting of a blackbody and an exponential cutoff power-law (BB+CPL) than the previously used \texttt{compTT} continuum model. 
This was supported by the spectral fitting of \edit1{{\it Beppo}SAX and {\it RXTE} observations} carried out in high and intermediate luminosity states of the pulsar.

\item
We identified  iron K$_\alpha$ and K$_\beta$ emission lines and newly detected 
K$_\alpha$ emission line at 7.49~keV from lowly ionized Ni atoms.

\item
Detection of iron K$_\alpha$, K$_\beta$ and K absorption edges indicated the degree 
of ionization of iron is less than 2 ($<$Fe I\hspace{-.1em}I\hspace{-.1em}I).

\item
The iron Ly$_{\alpha}$ line was clearly detected with an intensity comparable to that 
obtained from the {\it Chandra}/HETGS observation of the pulsar \citep{Paul2005}. 

\item
No cyclotron resonance scattering feature was detected in 30--110~keV range spectrum of the pulsar.

\item
The phase-sliced spectra can be well fitted by the BB+CPL continuum model. 
The parameters such as the power-law photon index and cutoff energy obtained 
from the phase-resolved spectroscopy showed a significant variation with pulse 
phase of the pulsar. However, the photoelectric absorption did not show any 
significant variation with pulse phase, including the dip interval.

\item
Clear spin phase modulation of the intensity of iron K$_{\alpha}$ emission line 
was detected with an amplitude of 7~\%, peaking at around 0.7 -- 1.1 phase.

\end{enumerate}

From the above results, we draw the following conclusions:

\hspace{1.5em} Although the \texttt{compTT} model can describe the broad-band spectra of 
GX~1+4, the parameters in the model are not sufficient to reproduce the emission spectrum 
due to Comptonization in the accretion column. A combination of blackbody and an exponential 
cutoff power-law can add another freedom in the model and can fit the observed broad-band 
spectrum better than the \texttt{compTT} model.

We derived the iron abundance in GX~1+4 to be $\sim$80~\% of the solar value as compared 
with the photoelectric absorption $N_{\rm H}$. Assuming both the iron and Ni fluorescent 
lines to be originated from same region, the abundance ratio [Ni/Fe] is calculated to be 
approximately two times larger than the solar value.

If the iron fluorescent line emitting region is homogeneous, the size of emission region 
is expected to be large in order to explain the observed low ionization state. 
However, a fine tuning is required to produce the line intensity modulation over 
pulsar phases. If we introduce an inhomogeneity in the matter distribution, 
smaller size of the fluorescent line emitting region can be accepted. 
This can also explain the line intensity modulation observed during 
{\it Suzaku} observation of the pulsar.

\acknowledgments
This work was  partially supported by the Ministry of Education, Culture, Sports, 
Science and Technology (MEXT), Grant-in-Aid for Science Research 25400237, and the 
MEXT Supported Program for the Strategic Research Foundation at Private Universities, 
2014-2018. This research was carried out by using data obtained from the Data Archive 
and Transmission System (DARTS), provided by Center for Science-satellite Operation and 
Data Archive (C-SODA) at ISAS/JAXA.


\begin{table*}[h]
\caption{Best-fit spectral parameters for emission lines and related structures obtained 
by fitting the phase-averaged spectrum of GX~1+4. \label{tab:PhaseAveLineParam}}
\begin{center}
\begin{tabular}{lllll}
\tableline\tableline
Model\tablenotemark{a}&Model\,A & Model\,A' & Model\,B & Model\,C \\
\tableline
$N_{\rm{H}}\,(10^{23}{\rm cm}^{-2})$	&1.30\,(fixed)	&1.30\,(fixed)	&1.30\,(fixed)	&1.30\,(fixed)	\\
$E_{\rm{Edge}}$\,(keV)	&$7.190^{+0.012}_{-0.014}$	&7.190\,(fixed)	&7.190\,(fixed)	&7.190\,(fixed)	\\
Max$\tau$	&$0.31\pm0.02$	&$0.30\pm0.02$	&$0.28\pm0.02$	&$0.31^{+0.01}_{-0.03}$	\\
$E_{\rm{Fe\,K}_\alpha}$\,(keV)	&$6.424\pm0.001$	&$6.425\pm0.001$	&$6.425^{+0.001}_{-0.002}$	&$6.424\pm0.001$	\\
$E_{\rm{Edge}}/E_{\rm{FeK}_\alpha}$	&$1.119\pm0.002$	&$1.119\pm0.002$	&$1.119\pm0.002$	&$1.119\pm0.002$	\\
$N_{\rm{Fe\,K}_\alpha}\,(10^{-3}\,\rm{photons\,s^{-1}\,cm^{-2}})$	&$1.39\pm0.02$	&$1.40\pm0.02$	&$1.29\pm0.02$	&$1.28\pm0.02$	\\
Eq.W$_{\rm{Fe\,K}_\alpha}$(eV)	&$152^{+3}_{-2}$	&$153\pm3$	&$153\pm3$	&$152^{+3}_{-2}$	\\
$E_{\rm{Fe\,K}_\beta}$\,(keV)\tablenotemark{b}	&7.086	&7.086	&7.087	&7.085	\\
$\sigma_{\rm{Fe\,K}_\beta}$\,(keV)	&0\,(fixed)	&$0.10^{+0.04}_{-0.03}$	&0\,(fixed)	&0\,(fixed)	\\
$N_{\rm{Fe\,K}_\beta}\,(10^{-4}\,\rm{photons\,s^{-1}\,cm^{-2}})$	&$1.5\pm0.2$	&$2.2^{+0.5}_{-0.4}$	&$1.3\pm0.3$	&$1.4\pm0.2$	\\
$N_{\rm{Fe\,K}_\beta}/N_{\rm{Fe\,K}_\alpha}$	&$0.11\pm0.02$	&$0.16\pm0.03$	&$0.10\pm0.02$	&$0.11\pm0.01$	\\
Eq.W$_{\rm{Fe\,K}_\beta}$\,(eV)	&$18^{+5}_{-6}$	&$25^{+5}_{-4}$	&$15\pm3$	&$17\pm3$	\\
$E_{\rm{Ni\,K}_\alpha}$\,(keV)	&$7.49\pm0.02$	&$7.50\pm0.02$	&$7.49\pm0.02$	&$7.48^{+0.02}_{-0.01}$	\\
$N_{\rm{Ni\,K}_\alpha}\,(10^{-4}\,\rm{photons\,s^{-1}\,cm^{-2}})$	&$1.7\pm0.3$	&$1.7\pm0.3$	&$1.3\pm0.2$	&$1.5\pm0.2$	\\
Eq.W$_{\rm{Ni\,K}_\alpha}$\,(eV)	&$17\pm5$	&$18\pm3$	&$16^{+3}_{-2}$	&$17\pm3$	\\
$E_{\rm{Fe\,Ly}_\alpha}$\,(keV)	&-	&-	&$6.98\pm0.06$	&-	\\
$N_{\rm{Fe\,Ly}_\alpha}\,(10^{-4}\,\rm{photons\,s^{-1}\,cm^{-2}})$	&-	&-	&$0.6\pm0.3$	&-	\\
Eq.W$_{\rm{Fe\,Ly}_\alpha}$(eV)	&-	&-	&$ 8\pm3$	&-	\\
$E_{\rm{Fe\,He}_\alpha}$\,(keV)	&-	&-	&-	&6.70\,(fixed)	\\
$N_{\rm{Fe\,He}_\alpha}\,(10^{-4}\,\rm{photons\,s^{-1}\,cm^{-2}})$	&-	&-	&-	&$<0.1$	\\
Eq.W$_{\rm{Fe\,He}_\alpha}$(eV)	&-	&-	&-	&$ <3.2$	\\
$\chi_{\nu}\,^{2}\,(d.o.f.)$	&1.10\,(966)	&1.09\,(965)	&1.09\,(965)	&1.11\,(965)	\\
\tableline
\end{tabular}
\end{center}
\tablecomments{Widths of iron K$_\alpha$, Ni\,K$_\alpha$, iron Ly$_\alpha$ and iron He$_\alpha$ were fixed to 0 eV. The errors given here are for 90~\% confidence limits.}
\tablenotetext{a}{
Model\,A: a power-law continuum multiplied by an absorption edge model along with three gaussians for iron K$_\alpha$ and K$_\beta$ lines and Ni\,K$_\alpha$ line. 
Model\,A': similar spectral components of Model\,A but the line width of iron K$_{\beta}$  remained a free parameter.
Model\,B,C: spectral models which are Model\,A with an additional gaussian component. Model\,B is included an additional gaussian for iron Ly$_\alpha$ line and Model\,C is included an additional gaussian for iron He$_\alpha$ line.
In any model, the photoelectric absorption is multiplied to only the gaussian components.}
\tablenotetext{b}{
Energies of iron K$_\beta$ line was set to be 1.103$\times$ of that of iron K$_\alpha$ line.}

\end{table*}

\begin{table*}[h]
\caption{Best-fit spectral parameters obtained by fitting the phase-sliced spectra of
{\it Suzaku} observation of GX~1+4. \label{tab:suzaku-phase-sliced}}
\begin{center}
\begin{tabular}{llllllll} \tableline\tableline
Interval\tablenotemark{a}& 1 \& 3 & 2 & 4 & 5 & 6 & 7 & 8 \\ 
\tableline
\multicolumn{8}{c}{\texttt{compTT} assuming disk geometry} \\
$N_{\rm{H}}(10^{23}\,\rm{cm^{-2}})$	&$1.13^{+0.04}_{-0.03}$	&$1.06^{+0.06}_{-0.04}$	&$1.16^{+0.04}_{-0.03}$	&$1.07^{+0.03}_{-0.02}$	&$1.09^{+0.03}_{-0.02}$	&$1.09^{+0.04}_{-0.02}$	&$1.07^{+0.03}_{-0.04}$	\\
$T_{\rm{o}}$\,(keV)	&$1.30^{+0.04}_{-0.05}$	&$1.66^{+0.05}_{-0.10}$	&$1.19^{+0.03}_{-0.05}$	&$1.40^{+0.02}_{-0.03}$	&$1.44^{+0.02}_{-0.03}$	&$1.38^{+0.02}_{-0.05}$	&$1.33^{+0.05}_{-0.04}$	\\
$T_{\rm{e}}$\,(keV)	&$ 9.4^{+0.3}_{-0.3}$	&$ 8.6^{+0.3}_{-0.4}$	&$11.3^{+0.3}_{-0.3}$	&$11.4^{+0.2}_{-0.2}$	&$11.3^{+0.1}_{-0.2}$	&$11.3^{+0.1}_{-0.2}$	&$10.8^{+0.4}_{-0.2}$	\\
$\tau$	&$ 4.3^{+0.1}_{-0.1}$	&$ 5.2^{+0.3}_{-0.2}$	&$ 3.8^{+0.1}_{-0.1}$	&$ 4.4^{+0.1}_{-0.1}$	&$ 4.6^{+0.1}_{-0.0}$	&$ 4.8^{+0.1}_{-0.0}$	&$ 4.3^{+0.1}_{-0.1}$	\\
$A_{\rm{c}}$\tablenotemark{b}	&$1.82^{+0.07}_{-0.06}$	&$1.17^{+0.05}_{-0.04}$	&$1.53^{+0.06}_{-0.05}$	&$1.78^{+0.03}_{-0.03}$	&$2.01^{+0.03}_{-0.02}$	&$1.98^{+0.04}_{-0.02}$	&$1.86^{+0.04}_{-0.07}$	\\
$\chi_{\nu}^{2}\,(d.o.f.)$	&1.17\,(261)	&1.23\,(261)	&1.30\,(261)	&1.86\,(261)	&1.82\,(261)	&1.50\,(261)	&1.41\,(261)	\\
\tableline
\multicolumn{8}{c}{\texttt{compTT} assuming sphere geometry} \\
$N_{\rm{H}}(10^{23}\,\rm{cm^{-2}})$	&$1.12^{+0.04}_{-0.03}$	&$1.07^{+0.05}_{-0.05}$	&$1.17^{+0.05}_{-0.03}$	&$1.06^{+0.03}_{-0.02}$	&$1.09^{+0.03}_{-0.02}$	&$1.09^{+0.03}_{-0.02}$	&$1.06^{+0.04}_{-0.03}$	\\
$T_{\rm{o}}$\,(keV)	&$1.31^{+0.03}_{-0.05}$	&$1.63^{+0.09}_{-0.08}$	&$1.19^{+0.02}_{-0.05}$	&$1.41^{+0.02}_{-0.04}$	&$1.44^{+0.02}_{-0.04}$	&$1.38^{+0.02}_{-0.04}$	&$1.34^{+0.04}_{-0.05}$	\\
$T_{\rm{e}}$\,(keV)	&$ 9.4^{+0.3}_{-0.3}$	&$ 8.6^{+0.4}_{-0.3}$	&$11.3^{+0.3}_{-0.3}$	&$11.4^{+0.2}_{-0.2}$	&$11.3^{+0.1}_{-0.2}$	&$11.3^{+0.1}_{-0.2}$	&$10.9^{+0.3}_{-0.3}$	\\
$\tau$	&$ 9.3^{+0.3}_{-0.2}$	&$11.2^{+0.4}_{-0.5}$	&$ 8.4^{+0.2}_{-0.1}$	&$ 9.5^{+0.2}_{-0.1}$	&$ 9.9^{+0.1}_{-0.1}$	&$10.4^{+0.2}_{-0.1}$	&$ 9.3^{+0.2}_{-0.2}$	\\
$A_{\rm{c}}$\tablenotemark{b}	&$1.83^{+0.07}_{-0.05}$	&$1.20^{+0.05}_{-0.05}$	&$1.55^{+0.07}_{-0.04}$	&$1.81^{+0.04}_{-0.03}$	&$2.04^{+0.03}_{-0.03}$	&$2.01^{+0.04}_{-0.02}$	&$1.87^{+0.06}_{-0.04}$	\\
$\chi_{\nu}^{2}\,(d.o.f.)$	&1.17\,(261)	&1.24\,(261)	&1.30\,(261)	&1.86\,(261)	&1.82\,(261)	&1.50\,(261)	&1.41\,(261)	\\
\tableline
\multicolumn{8}{c}{BB+CPL\tablenotemark{c}} \\
$N_{\rm{H}}(10^{23}\,\rm{cm^{-2}})$	&$1.38^{+0.07}_{-0.06}$	&$1.31^{+0.09}_{-0.07}$	&$1.35^{+0.05}_{-0.05}$	&$1.32^{+0.03}_{-0.04}$	&$1.29^{+0.03}_{-0.03}$	&$1.25^{+0.04}_{-0.02}$	&$1.33^{+0.04}_{-0.07}$	\\
$kT_{\rm{BB}}$\,(keV)	&$1.6\pm0.2$	&$2.1^{+0.8}_{-0.3}$	&$1.49^{+0.07}_{-0.08}$	&$1.67^{+0.09}_{-0.08}$	&$1.75^{+0.06}_{-0.07}$	&$1.66^{+0.05}_{-0.08}$	&$1.6\pm0.2$	\\
$R$\,(km)\tablenotemark{d}	&$0.53^{+0.44}_{-0.46}$	&$0.27^{+0.27}_{-0.24}$	&$0.81^{+0.38}_{-0.39}$	&$0.58^{+0.31}_{-0.30}$	&$0.66^{+0.27}_{-0.28}$	&$0.78^{+0.28}_{-0.27}$	&$0.56^{+0.39}_{-0.39}$	\\
$\Gamma$	&$0.68^{+0.16}_{-0.18}$	&$0.11^{+0.20}_{-0.24}$	&$0.71^{+0.12}_{-0.13}$	&$0.54^{+0.07}_{-0.09}$	&$0.31^{+0.08}_{-0.08}$	&$0.16^{+0.12}_{-0.07}$	&$0.66^{+0.10}_{-0.16}$	\\
$E_{\rm{cutoff}}$\,(keV)	&$19.8^{+2.5}_{-2.3}$	&$14.5^{+1.8}_{-1.6}$	&$24.3^{+2.6}_{-2.3}$	&$24.2^{+1.3}_{-1.5}$	&$21.2^{+1.1}_{-1.0}$	&$20.0^{+1.7}_{-0.8}$	&$24.2^{+2.0}_{-2.6}$	\\
$\chi_{\nu}^{2}\,(d.o.f.)$	&1.02\,(260)	&1.15\,(260)	&1.14\,(260)	&1.19\,(260)	&1.28\,(260)	&1.10\,(260)	&1.14\,(260)	\\
\tableline
\multicolumn{8}{c}{lines and related structures\tablenotemark{e}} \\
$N_{\rm{H}}\,(10^{22}{\rm cm}^{-2})$	&1.38(fixed)	&1.31(fixed)	&1.35(fixed)	&1.32(fixed)	&1.29(fixed)	&1.25(fixed)	&1.33(fixed)	\\
Max$\tau$	&$0.34^{+0.05}_{-0.05}$	&$0.39^{+0.04}_{-0.05}$	&$0.31^{+0.04}_{-0.04}$	&$0.30^{+0.03}_{-0.04}$	&$0.27^{+0.04}_{-0.03}$	&$0.34^{+0.04}_{-0.04}$	&$0.34^{+0.05}_{-0.05}$	\\
$E_{\rm{FeK}\alpha}$\,(keV)	&$6.428^{+0.004}_{-0.006}$	&$6.425^{+0.003}_{-0.005}$	&$6.426^{+0.004}_{-0.004}$	&$6.424^{+0.004}_{-0.002}$	&$6.423^{+0.003}_{-0.003}$	&$6.425^{+0.003}_{-0.002}$	&$6.422^{+0.004}_{-0.004}$	\\
$N_{\rm{FeK}\alpha}$\tablenotemark{f}	&$1.44^{+0.06}_{-0.09}$	&$1.45^{+0.06}_{-0.06}$	&$1.31^{+0.06}_{-0.06}$	&$1.31^{+0.04}_{-0.06}$	&$1.37^{+0.05}_{-0.05}$	&$1.44^{+0.06}_{-0.06}$	&$1.52^{+0.08}_{-0.08}$	\\
Eq.W$_{\rm{FeK}\alpha}$\,(eV)	&$157.9^{+17.2}_{-8.7}$	&$301.9^{+138.0}_{-31.3}$	&$145.2^{+9.1}_{-6.2}$	&$136.3^{+7.3}_{-4.9}$	&$132.8^{+6.9}_{-5.6}$	&$152.3^{+15.3}_{-6.1}$	&$153.9^{+13.1}_{-6.6}$	\\
$N_{\rm{FeK}\beta}$\tablenotemark{f}	&$0.13^{+0.06}_{-0.08}$	&$0.20^{+0.05}_{-0.03}$	&$0.17^{+0.05}_{-0.05}$	&$0.12^{+0.05}_{-0.05}$	&$0.10^{+0.06}_{-0.05}$	&$0.15^{+0.06}_{-0.05}$	&$0.26^{+0.07}_{-0.08}$	\\
Eq.W$_{\rm{FeK}\beta}$\,(eV)	&$15.1^{+16.9}_{-15.1}$	&$41.3^{+36.5}_{-18.6}$	&$20.5^{+15.0}_{-14.7}$	&$13.4^{+11.3}_{-9.1}$	&$10.2^{+ 7.1}_{-10.2}$	&$16.7^{+12.2}_{-9.7}$	&$28.7^{+13.0}_{-17.3}$	\\
$E_{\rm{NiK}\alpha}$\,(keV)	&$7.45^{+0.04}_{-0.04}$	&$7.51^{+0.03}_{-0.03}$	&$7.54^{+0.04}_{-0.05}$	&$7.49^{+0.03}_{-0.04}$	&$7.45^{+0.12}_{-0.09}$	&$7.48^{+0.04}_{-0.04}$	&$7.46^{+0.04}_{-0.04}$	\\
$N_{\rm{NiK}\alpha}$\tablenotemark{f}	&$0.26^{+0.09}_{-0.10}$	&$0.19^{+0.07}_{-0.07}$	&$0.17^{+0.07}_{-0.07}$	&$0.20^{+0.06}_{-0.07}$	&$0.11^{+0.07}_{-0.07}$	&$0.22^{+0.07}_{-0.08}$	&$0.27^{+0.09}_{-0.09}$	\\
Eq.W$_{\rm{Ni}}$\,(eV)	&$26.5^{+18.8}_{-14.4}$	&$34.2^{+39.8}_{-19.5}$	&$18.4^{+11.5}_{-14.1}$	&$19.3^{+11.8}_{-8.9}$	&$10.0^{+11.5}_{-10.0}$	&$21.0^{+11.1}_{-9.6}$	&$25.4^{+16.2}_{-11.8}$	\\
$\chi_{\nu}^{2}(d.o.f)$	&1.07(92)	&1.05(92)	&1.20(92)	&1.12(92)	&1.52(92)	&1.27(92)	&1.24(92)	\\
\tableline
\end{tabular}
\end{center}
\tablecomments{The errors given here are for 90\% confidence limits.}
\tablenotetext{a}{The interval 1, interval 2, interval 3, interval 4, interval 5, interval 6, interval 7 and interval 8 correspond to 0.90--0.95, 0.95--0.05, 0.05--0.10, 0.10--0.25, 0.25--0.45, 0.45--0.65, 0.65--0.80 and 0.80--0.90 phase ranges, respectively.}
\tablenotetext{b}{Normalization parameter for the \texttt{compTT} model component ($\times 10^{-2}$).}
\tablenotetext{c}{BB and CPL represent blackbody, exponential cutoff power-law, respectively.}
\tablenotetext{d}{Blackbody radius assuming a distance of 4.3~kpc.}
\tablenotetext{e}{
The model is represented by a power-law continuum multiplied by an absorption edge model along with three gaussians for iron K$_\alpha$ and K$_\beta$ lines and Ni\,K$_\alpha$ line.
Widths of iron K$_\alpha$, iron K$_\beta$ and Ni\,K$_\alpha$ are fixed to 0 eV in the fitting.
Energies of  iron K$_\beta$ line and the edge are set to be 1.103$\times$ and 1.119$\times$ of that of iron K$_\alpha$ line, respectively.}
\tablenotetext{f}{In unit of $10^{-3}\,\mathrm{photons\,s^{-1}\,cm^{-2}}$.}
\end{table*}

\begin{table*}[h]
\caption{Log of {\it Beppo}SAX and {\it RXTE} observations of GX~1+4 and spectral parameters obtained from fitting data with three different models \label{tab:xslog}}
\begin{center}
\begin{tabular}{llllllllll} \tableline\tableline
Observatory&{\it Beppo}SAX&{\it Beppo}SAX&{\it RXTE}&{\it RXTE}&{\it RXTE}&{\it RXTE} \\
Observation&S-I&S-I\hspace{-.1em}I&X-I &X-I\hspace{-.1em}I &X-I\hspace{-.1em}I\hspace{-.1em}I&X-I\hspace{-.1em}V  \\
ID	&2020500100 &2020500200&10133-02-01-000 &10133-02-01-001&10104-02-01-00&70065-01-01-00 \\
Date\,(yyyy-mm-dd)& 1996-08-18 &1997-05-25&1996-02-12	&1996-02-12	&1996-02-17		&2002-04-26 	\\
Exposure\,(ks)	& 77.2&62.6&26.5	&21.0	&18.0	&20.6 \\ 
\tableline
\multicolumn{7}{c}{\texttt{compTT} assuming disk geometry} \\
$N_{\rm{H}}(10^{22}\,\rm{cm^{-2}})$	&$18.8^{+1.0}_{-1.0}$	&$ 0.9^{+0.1}_{-0.1}$	&$5.7\pm0.1$	&$6.3\pm0.1$	&$5.4\pm0.1$	&$0.6\pm0.2$	\\
$T_{\rm{o}}$\,(keV)	&$1.65^{+0.05}_{-0.05}$	&$1.35^{+0.03}_{-0.03}$	&$1.54\pm0.01$	&$1.62\pm0.01$	&$1.61\pm0.01$	&$1.46\pm0.01$	\\
$T_{\rm{e}}$\,(keV)	&$13.4^{+0.5}_{-0.4}$	&$14.1^{+1.7}_{-1.3}$	&$11.0\pm0.1$	&$9.5\pm0.1$	&$10.2\pm0.1$	&$11.8\pm0.1$	\\
$\tau$	&$ 2.8^{+0.1}_{-0.1}$	&$ 2.5^{+0.3}_{-0.3}$	&$4.05\pm0.02$	&$4.33\pm0.03$	&$4.74\pm0.02$	&$3.58\pm0.03$	\\
$A_{\rm{c}}$\tablenotemark{$a$}	&$7.7^{+0.4}_{-0.4}$	&$4.5^{+0.6}_{-0.6}$	&$24.2\pm0.2$	&$21.7\pm0.2$	&$39.3\pm0.2$	&$27.4\pm0.3$	\\
$\chi_{\nu}^{2}\,(d.o.f.)$	&1.10\,(261)	&1.09\,(363)	&3.11\,(191)	&5.00\,(191)	&5.02\,(126)	&1.85\,(174)	\\
\tableline
\multicolumn{7}{c}{\texttt{compTT} assuming sphere geometry} \\
$N_{\rm{H}}(10^{22}\,\rm{cm^{-2}})$	&$18.8^{+1.4}_{-1.0}$	&$ 0.9^{+0.1}_{-0.1}$	&$5.7\pm0.1$	&$6.3\pm0.1$	&$5.4\pm0.1$	&$0.6\pm0.2$	\\
$T_{\rm{o}}$\,(keV)	&$1.65^{+0.05}_{-0.05}$	&$1.35^{+0.03}_{-0.03}$	&$1.54\pm0.01$	&$1.62\pm0.01$	&$1.61\pm0.01$	&$1.46\pm0.01$	\\
$T_{\rm{e}}$\,(keV)	&$13.4^{+0.5}_{-0.4}$	&$14.1^{+1.6}_{-1.3}$	&$11.0\pm0.1$	&$9.5\pm0.1$	&$10.2\pm0.1$	&$11.8\pm0.1$	\\
$\tau$	&$ 6.3^{+0.2}_{-0.2}$	&$ 5.7^{+0.6}_{-0.6}$	&$4.05\pm0.02$	&$4.33\pm0.03$	&$4.74\pm0.02$	&$3.58\pm0.03$	\\
$A_{\rm{c}}$\tablenotemark{$a$}	&$7.7^{+0.4}_{-0.4}$	&$4.5^{+0.6}_{-0.6}$	&$24.2\pm0.2$	&$21.7\pm0.2$	&$39.3\pm0.2$	&$27.4\pm0.3$	\\
$\chi_{\nu}^{2}\,(d.o.f.)$	&1.10\,(261)	&1.09\,(363)	&3.11\,(191)	&5.00\,(191)	&5.02\,(126)	&1.85\,(174)	\\
\tableline
\multicolumn{7}{c}{BB+CPL\tablenotemark{$b$}} \\
$N_{\rm{H}}(10^{22}\,\rm{cm^{-2}})$	&$23.46^{+1.40}_{-1.45}$	&$2.39^{+0.20}_{-0.21}$	&$8.7\pm0.2$	&$10.5\pm0.2$	&$8.2\pm0.2$	&$3.6\pm0.2$	\\
$kT_{\rm{BB}}$\,(keV)	&$2.4^{+1.6}_{-0.3}$	&$2.0^{+0.1}_{-0.1}$	&$2.09\pm0.03$	&$3.18\pm0.09$	&$2.00\pm0.03$	&$2.01\pm0.03$	\\
$R_{\rm{BB}}$\,(km)\tablenotemark{$c$}	&$0.23^{+0.28}_{-0.22}$	&$2.16^{+0.72}_{-0.68}$	&$0.5\pm0.2$	&$0.2\pm0.1$	&$0.6\pm0.2$	&$0.7\pm0.2$	\\
$\Gamma$	&$0.80^{+0.09}_{-0.13}$	&$0.89^{+0.09}_{-0.11}$	&$0.60\pm0.03$	&$0.79\pm0.03$	&$0.33\pm0.03$	&$0.77\pm0.03$	\\
$E_{\rm{cutoff}}$\,(keV)	&$24.5^{+1.6}_{-1.8}$	&$25.3^{+2.8}_{-2.5}$	&$23.6\pm0.5$	&$27.8\pm1.1$	&$19.6\pm0.4$	&$26.4\pm0.9$	\\
$\chi_{\nu}^{2}\,(d.o.f.)$	&1.00\,(260)	&1.00\,(362)	&1.51\,(190)	&1.90\,(190)	&1.49\,(125)	&1.06\,(173)	\\
\tableline
\end{tabular}
\end{center}
\tablecomments{The errors given here are for 90\% confidence limits in {\it Beppo}SAX data and 1$\sigma$ confidence limits in {\it RXTE} data.}
\tablenotetext{$a$}{Normalization parameter for the \texttt{compTT} model component ($\times 10^{-3}$).}
\tablenotetext{$b$}{BB and CPL represent blackbody, exponential cutoff power-law, respectively.}
\tablenotetext{$c$}{Blackbody radius assuming a distance of 4.3~kpc.}
\end{table*}

\newpage
\appendix
\section{Correction for SCF effect}
\begin{figure*}[h]
\gridline{\fig{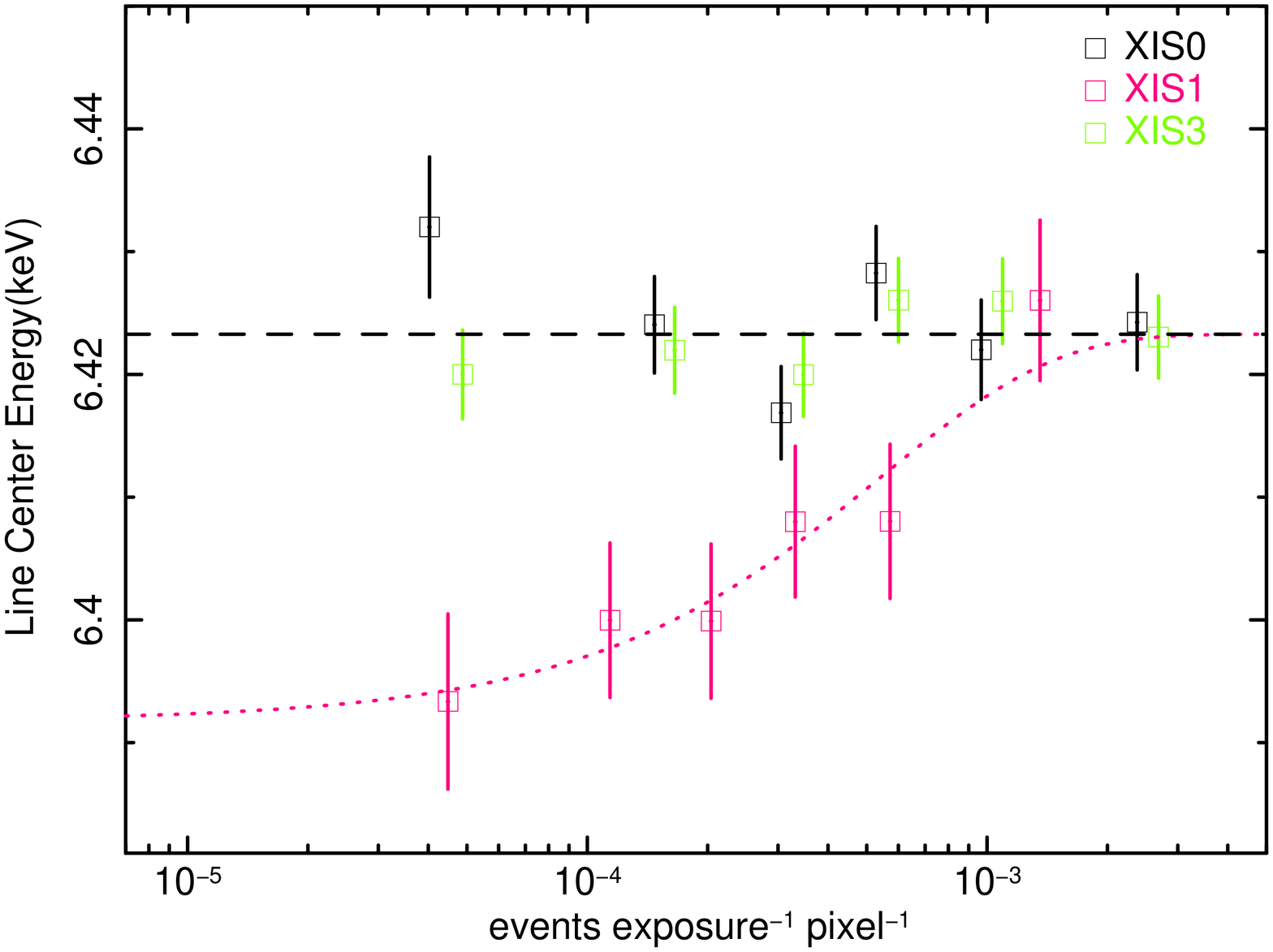}{0.4\textwidth}{(a)}
	     \fig{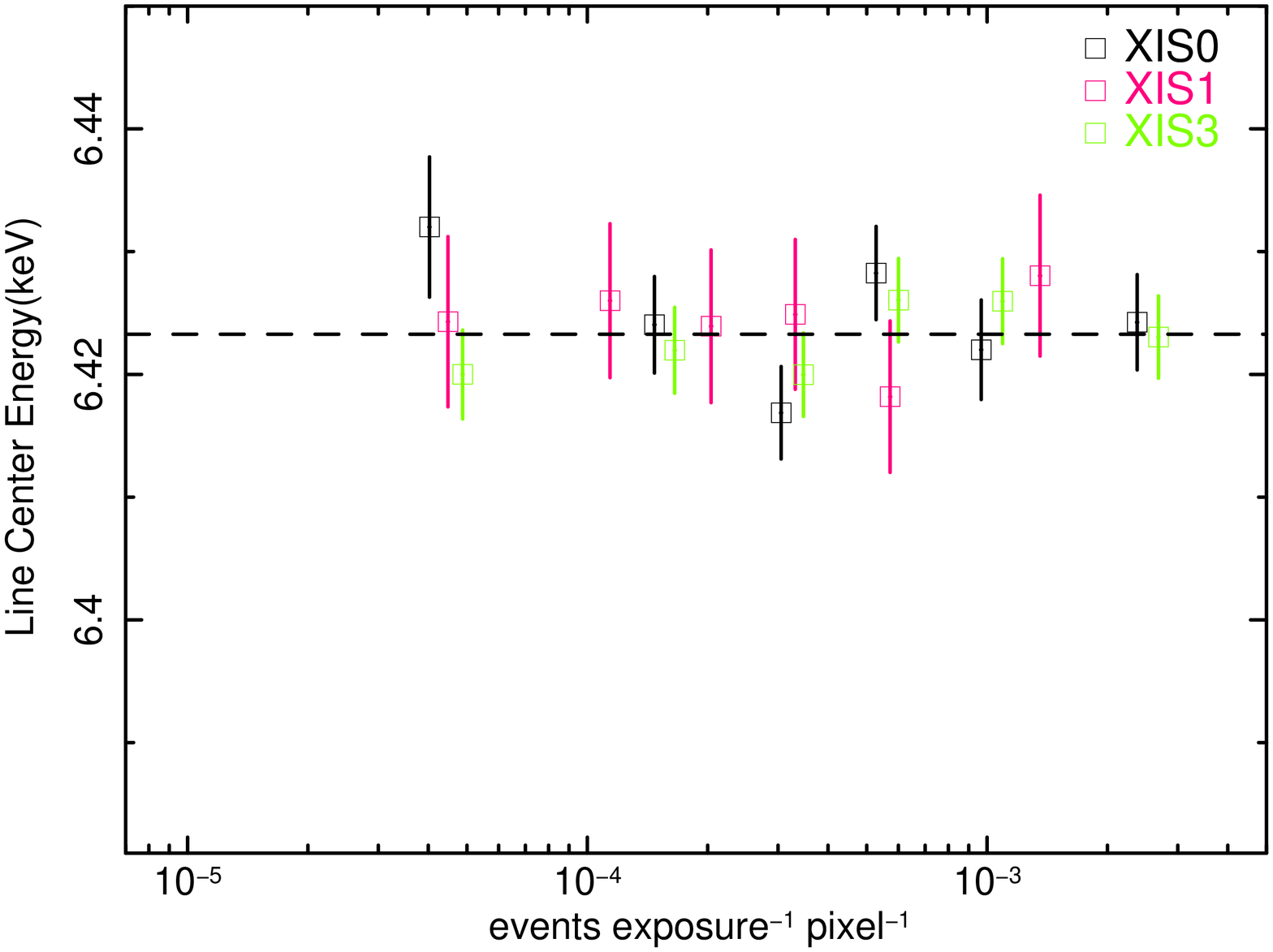}{0.4\textwidth}{(b)}}
	\caption{The central energy of the iron K$_\alpha$ emission line as a function of event density. Panel\,(a) shows before correction of SCF effect of XIS~1 data and panel\,(b) shows after the correction.\newline} 
	\label{fig:SCF}
\end{figure*}

The Self Charge Filling (SCF) effect of {\it Suzaku} data was first pointed out by 
\citet{Todoroki2012} and consequently they proposed a method to correct the data for
this effect. As suggested, we divided the source extraction region of 3' radius into 
six regions. Energy spectrum was extracted from each of the regions. The prominent 
iron K $_\alpha$ emission line in each spectrum was fitted by a Gaussian function 
and the line energy was derived from fitting. The central energies obtained from 
XIS~0, XIS~1, and XIS~3 data were plotted as a function of the event density (in 
unit of ``events\,exposure$^{-1}$\,pixel$^{-1}$'') and shown in Figure~\ref{fig:SCF}.  
A clear dependence of the central energy on the event density was found for XIS-1 data, 
but no apparent dependency was found in data obtained from XIS-0 and XIS-3.
Then we fitted the central energy of XIS~1 with the function of
\begin{equation}
E = E_0+(E_t-E_0)[1- \exp{(-\epsilon x)}] = E_t [1-C \exp{(-\epsilon x)}]
\end{equation}
\noindent
where $x$ is the event density, $E_t$ is the expected true energy, $\epsilon$ 
is the amount of the SCF effect, $E_0$ is the energy in the low event density 
limit and $C= (E_t-E_0)/E_t$. The best fit curve is shown in Figure~\ref{fig:SCF} 
with the dotted line, and the best fit parameters are listed in Table~\ref{tab:SCF}.

The energy scale of the spectra obtained from the six regions of XIS~1 was corrected 
according to the parameters obtained (shown in Table~\ref{tab:SCF}). The central energy 
of iron K$_\alpha$ emission line obtained from the corrected spectra of XIS~1 are plotted 
in Figure~\ref{fig:SCF}, as well as the data of XIS~0 and XIS~3.

\begin{table*}
	\begin{center}
	\caption{Best fit parameters of SCF effect for XIS~1.\label{tab:SCF}}
	\begin{tabular}{lll}
		\tableline\tableline
		Parameter & Unit & Value \\
		\tableline
		$E_t$ & keV & 6.42 \\
		$E_0$ & keV & 6.39$\pm0.01$ \\
		$\epsilon $ &  - & 1800$^{+1800}_{-1000} $\\
		\tableline
	\end{tabular}
	\end{center}
\end{table*}

%




\end{document}